\documentclass[12pt]{article}
\usepackage{latexsym}
\usepackage{epsfig,amssymb,euscript,slashed}
\usepackage{amsmath}
\usepackage[nosort]{cite} 
\usepackage{array,calc,epsfig}
\usepackage{bbm}
\usepackage{fancybox}
\usepackage[linktocpage]{hyperref}

\numberwithin{equation}{section} 
\usepackage[]{graphicx}

\usepackage{color}

\begin{document}
\font\cmss=cmss10 \font\cmsss=cmss10 at 7pt

\begin{flushright}{\scriptsize DFPD-17-TH-07 \\  
\scriptsize QMUL-PH-17-08}
\end{flushright}
\hfill
\vspace{18pt}
\begin{center}
{\Large 
\textbf{Holographic 4-point correlators with heavy states}
}
\end{center}

\vspace{8pt}
\begin{center}
{\textsl{Andrea Galliani$^{\,a, b}$, Stefano Giusto$^{\,a, b}$ and Rodolfo Russo$^{\,c}$}}

\vspace{1cm}

\textit{\small ${}^a$ Dipartimento di Fisica ed Astronomia ``Galileo Galilei",  Universit\`a di Padova,\\Via Marzolo 8, 35131 Padova, Italy} \\  \vspace{6pt}

\textit{\small ${}^b$ I.N.F.N. Sezione di Padova,
Via Marzolo 8, 35131 Padova, Italy}\\
\vspace{6pt}

\textit{\small ${}^c$ Centre for Research in String Theory, School of Physics and Astronomy\\
Queen Mary University of London,
Mile End Road, London, E1 4NS,
United Kingdom}\\
\vspace{6pt}

\end{center}

\vspace{12pt}

\begin{center}
\textbf{Abstract}
\end{center}

\vspace{4pt} {\small
\noindent 
The AdS/CFT duality maps supersymmetric heavy operators with conformal dimension of the order of the central charge to asymptotically AdS supergravity solutions. We show that by studying the quadratic fluctuations around such backgrounds it is possible to derive the 4-point correlators of two light and two heavy states in the supergravity approximation. We provide an explicit example in the AdS$_3$ setup relevant for the duality with the D1-D5 CFT. Contrary to previously studied examples, the supergravity correlator derived in this work differs from the result obtained at the CFT orbifold point.  Our method bypasses the difficulties of applying the standard Witten's diagrams approach to correlators with operators of large conformal dimension and also avoids some technical steps that have made the computation of dynamical 4-point correlators in the AdS$_3$/CFT$_2$ context unfeasible until now.}

\vspace{1cm}


\thispagestyle{empty}

\vfill
\vskip 5.mm
\hrule width 5.cm
\vskip 2.mm
{
\noindent  {\scriptsize e-mails:  {\tt andrea.galliani@pd.infn.it, stefano.giusto@pd.infn.it, r.russo@qmul.ac.uk} }
}

\setcounter{footnote}{0}
\setcounter{page}{0}

\newpage

\section{Introduction}

The analysis of 4-point correlators has been one of the key ways to study the AdS$_{d+1}$/CFT$_d$ duality~\cite{Maldacena:1997re,Gubser:1998bc,Witten:1998qj} from its very early days. While 3-point functions of operators belonging to supersymmetric chiral multiplets do not depend on the coupling constants,\footnote{The general proof of this non-renormalization theorem was given in \cite{Baggio:2012rr}, and in \cite{Kanitscheider:2006zf,Kanitscheider:2007wq,Giusto:2015dfa} it was exploited to match states at weak and string coupling in the particular set-up relevant for this article.} 4-point correlators of the same operators are generically non-protected quantities that capture interesting dynamical features of the theory. Most of the work in this area has been so far focused on the $d>2$ case and, in particular, on correlators of single trace operators for the four dimensional ${\cal N}=4$ SYM theory. On the bulk side, the standard approach for evaluating these correlators is to use Witten's diagrams, but other approaches are being developed which bypass part of the supergravity analysis, see~\cite{Rastelli:2016nze} and references therein also for an overview of the results obtained with the traditional approach.

In this article we extend these results in two ways. Firstly, we compute correlators at strong coupling in a 2D CFT, where the conventional Witten-diagram method faces some technical difficulties, because the quartic couplings of the 3D supergravity Lagrangian have not been worked out and the diagrams representing vector and graviton exchange are naively divergent in AdS$_3$ \cite{DHoker:1999mqo}. Hence, dynamical correlators have not been computed  in the AdS$_3$/CFT$_2$ setting until now. Secondly, we consider correlators that contain ``heavy''  operators $O_H$, whose conformal dimensions scale as the central charge $c$ of the theory when $c\to\infty$; the $O_H$'s  are typically multiparticle operators with large quantum numbers. We will refer to ordinary single particle operators $O_L$, with dimensions of order one in the same limit, as ``light''. In analogy with the case of ${\cal N}=4$ SYM we also use the nomenclature single or multitrace operators to indicate single or multiparticle states. We concentrate in particular on 4-point functions with two heavy and two light operators, which we will denote as HHLL. This class of correlators has recently been studied both from the CFT~\cite{Balasubramanian:2005qu,Lunin:2012gz,Galliani:2016cai,Balasubramanian:2016ids} and the bulk point of view~\cite{Fitzpatrick:2015zha,Hijano:2015zsa,Hijano:2015qja,daCunha:2016crm}). They are particularly interesting in the study of black holes in the AdS/CFT context as the heavy operators represent the microstates that provide a statistical interpretation of the Bekenstein-Hawking entropy. 

It should be noted that a different type of heavy operators were considered before: these are non-protected semiclassical stringy operators of dimension of order $c^{1/4}$: 3-point functions of the HHL type were computed in \cite{Zarembo:2010rr,Costa:2010rz}, and the extension to 4-point HHLL correlators was worked out in~\cite{Buchbinder:2010ek}. There is a substantial difference between the two types of heavy operators: when the dimension is $O(c^{1/4})$, $O_H$ is dual to a semiclassical string state in AdS, while operators with $O(c)$ dimension source new asymptotically AdS geometries. Since neither type of heavy operators is dual to a simple supergravity mode, it does not seem to be easy to use the standard Witten's diagrams approaches mentioned above to derive the HHLL correlators. For the stringy operators of~\cite{Zarembo:2010rr,Costa:2010rz,Buchbinder:2010ek}, one could use the worldsheet theory dual to the semiclassical states to compute holographically the correlators, but this method is not applicable for the correlators of interest here. Another powerful technique is integrability, which has been a very successful approach in studying the spectrum and the correlation functions of holographic superconformal theories, see~\cite{Sfondrini:2014via} for a review in the AdS$_3$/CFT$_2$ context; however the correlators we are interested in are outside the integrable sector of the theory since they involve operators that have a conformal dimension of the order of the central charge. As in the case of the AdS$_5$/CFT$_4$ SYM duality, we do not expect integrability to help in understanding the properties of the heavy states representing the microstates of a black hole. In this paper we develop a new holographic approach to the problem of calculating correlators with heavy operators of dimension of order $c$, which exploits the supergravity solution dual to the heavy states. We restrict here to supersymmetric operators. 

We focus on the $d=2$ case that relates a $1+1$-dimensional CFT with $(4,4)$ supersymmetries, known as the D1-D5 CFT, to type IIB string theory on AdS$_3 \times S^3 \times {\cal M}$, where the four dimensional compact space $\mathcal{M}$ can be either $T^4$ or K3. This background geometry arises in the near-horizon limit of a stack of $n_1$ D1 and $n_5$ D5-branes wrapped on a common $S^1$ and (for the D5's) on ${\cal M}$. The dual CFT has central charge $c=6 N$ with $N = n_1 n_5$ and its moduli space contains a free point where the CFT can be described simply as an orbifold with target space ${\cal M}^N/S_N$. The states in the Ramond-Ramond (RR) sector of the CFT are heavy since their holomorphic and anti-holomorphic conformal dimensions are at least $c/24$. The supersymmetric states in this sector are relevant for the statistical interpretation of the entropy for a black hole in type IIB supergravity on $S^1\times {\cal M}$~\cite{Sen:1995in,Strominger:1996sh}. 

In particular we study correlators of the form
\begin{equation}
  \label{eq:corrHHLLint}
  \langle O_H(z_1,\bar{z}_1) \bar{O}_H(z_2,\bar{z}_2)
 O_L(z_3,\bar{z}_3) \bar{O}_L(z_4,\bar{z}_4) \rangle\;,
\end{equation}
where the $O_L$ is a chiral primary operator of dimension $h=\bar{h}=1/2$ and $O_H$ is the RR ground state that was studied in~\cite{Kanitscheider:2007wq,Giusto:2015dfa} in the precision holography analysis of the 3-point functions~\cite{Kanitscheider:2006zf}. At the orbifold point the free field realisation of the CFT can be used to rewrite~\eqref{eq:corrHHLLint} as a combination of free correlators determined by the single trace constituents of the heavy state. On the bulk side, the heavy state we consider is described in terms of a supersymmetric 10-dimensional type IIB geometry that has D1 and D5 charges determined by $n_1$ and $n_5$~\cite{Lunin:2001jy,Lunin:2002iz,Kanitscheider:2007wq}. In the decoupling limit, this solution reduces to a 6-dimensional solution (after a standard KK reduction on ${\cal M}$) that is asymptotically AdS$_3 \times S^3$. The correlator~\eqref{eq:corrHHLLint} is captured by studying the quadratic perturbation of this geometry where we add the supergravity field dual to $O_L$ with the appropriate boundary condition on the AdS boundary. A class of bulk correlators of this type was studied in~\cite{Galliani:2016cai}: the heavy state chosen there was composed of equal single trace constituents and the dual geometry could be factorised (in a particular coordinate system) as the product of an orbifold of AdS$_3$ times $S^3$. Here we consider a more generic heavy state composed of two different types of single trace constituents. In the orbifold CFT, each constituent is characterised by a winding number, which specifies the twist sector, and by the $SU(2)_L \times SU(2)_R$ R-charge. In the heavy state of~\cite{Galliani:2016cai} each constituent had winding one and maximal R-charge; here we add a new type of constituent, with the same winding number but vanishing R-charge. Since the total winding number is fixed in terms of the CFT central charge, the heavy state depends on a single parameter (related to $B$ in~\eqref{eq:SKs}) which controls the relative number of the two types of single trace constituents; the state preserves the same (eight) supercharges for any value of $B$ and it reduces to the state of~\cite{Galliani:2016cai} for $B=0$. When $B\not=0$, the dual geometry is a more complicated space which cannot be factorised in AdS$_3$ and $S^3$ factors: of course, the majority of the bulk microstates are of this type.

The gravity computation of the correlator of two of these heavy states and two light states involves a perturbation around the D1-D5 non-factorised geometry dual to the heavy state; the analysis of this perturbation requires some non-trivial calculations. First one needs to find the linearized equations of motion for the perturbation around a background, which, apart from being non-factorized, displays non-trivial values for all type IIB fields. This requires generalising the analysis of \cite{Deger:1998nm}, which applies to perturbations around AdS$_3\times S^3$. Then one has to reduce the six-dimensional linearized equations to a system of three-dimensional equations (in the asymptotically AdS$_3$ part of the space) describing the field dual to the light operator $O_L$. This step is obviously complicated by the non-factorised form of the background. Here we simplify this task by performing a perturbative expansion that is motivated by the result of the dual correlator at the orbifold point. The orbifold result~\eqref{eq:Gsk} has a simple polynomial dependence on the parameter $B$, which, on the gravity side, controls the deviation from the factorised geometry considered in~\cite{Galliani:2016cai}. This suggests to expand the gravity equations in this parameter: we will keep here only the first non-trivial order. In this way we can organise the computation using the basis of spherical harmonics on the $S^3$ of the factorised $B=0$ background and obtain a set of solvable three-dimensional bulk equations. This procedure, though perturbative in $B$, allows us to keep the exact dependence of the correlator on the cross-ratio $z$ obtained from the points points $z_i$ in~\eqref{eq:corrHHLLint}: in particular we have access to the regime in which the light operators are maximally far and the gravity correlator probes the D1-D5 geometry arbitrarily deep in the interior. 

We will not provide a detailed comparison between the free orbifold and supergravity correlators, but some simple features are immediately visible. First the correlator analysed is not protected, contrary to what happens to the $B=0$ case studied in~\cite{Galliani:2016cai}. This fact is not surprising, and can be understood already from the analysis of the single trace operators exchanged between the two light operators: While, as was shown in ~\cite{Galliani:2016cai}, the only operators exchanged at $B=0$ are protected operators consisting of chiral-algebra descendants of the identity, when $B\not=0$ the correlator at the orbifold point receives contributions also from the exchange of non-protected operators, which are expected to lift in the gravity limit. Hence the $O(B^2)$ part of the correlator is sensitive to dynamical effects that depend on the CFT moduli. It also receives contributions from the exchange of multiparticle operators whose dimension gets corrections at subleading orders in $1/N$. The latter feature is generic for the case of non-protected correlators in AdS/CFT and the logarithms of the correlator captures the anomalous dimension~\cite{D'Hoker:1999jp,Arutyunov:2000ku}. We leave the analysis of the operators exchanged in the various OPE channels to a future work. This is clearly of great interest as it has the potential to shed light on the mechanism by which the information of the heavy state is encoded in the gravity correlator. Another interesting question is whether from the geometries discussed in this paper it is possible to extract a standard light 4-point correlator in the NS sector. This would represent a significant technical advance, since, as it was mentioned above, none of these correlators could be computed with conventional Witten-diagrams techniques in AdS$_3$.  Notice however that the small $B$ approximation discussed above is a double scaling limit where $N$ is taken to be large with $b^2\sim B^2/N$ small but constant. Thus, even after performing a spectral flow transformation, the heavy state we consider remains a multiparticle operator with a number of constituents (proportional to $B^2$) that is of order $N$. Because of this, our approach cannot at present make direct contact with standard light 4-point correlators. It would also be interesting to study quantitatively the connection between the standard HHLL scaling and the correlators where the two light operators are substituted by two ``perturbatively'' heavy operators (whose conformal dimension scales as $\alpha c$ with $\alpha \ll 1$). This problem was studied very recently in the context of the Liouville theory~\cite{Balasubramanian:2017fan}.

The paper is organised as follows. In the next section we introduce the CFT operators that enter the 4-point correlator and give its expression at the orbifold point. Section \ref{sec:gravity} contains the gravity computations: We first describe the background geometries dual to RR ground states and derive the linearized equations satisfied by the perturbation dual to the light operator $O_L$ around such backgrounds. We then specialise to one particular family of RR ground states, which depends on a parameter $B$, and solve the perturbation equations in this background at the first non-trivial order in $B$. From the solution for the perturbation, with appropriate boundary conditions, we extract the 4-point correlator at strong coupling. In Section \ref{sec:outl} we perform some preliminary consistency checks on the holographic result and conjecture some of its consequences for the information loss paradox. The details of the computation of the free correlator are presented in Appendix \ref{sec:free-corr}. In Appendix \ref{sec:d1d5metric} we collect the supergravity solution dual to the heavy operator $O_H$ and its $B$-expansion. Some properties of $S^3$ spherical harmonics useful for the gravity computation are listed in Appendix \ref{sec:harmonics}. Appendix \ref{sec:cftconv} provides a short summary of the $D$-integrals that are ubiquitous in the supergravity results for the 4-point correlators and contains some useful identities they satisfy. Finally the steps needed to rewrite the gravity correlator in terms of $D$-integrals are detailed in Appendix~\ref{sec:red2dintegrals}. In Appendix~\ref{sec:Opmap} we provide the result for a correlator that has a different choice of the light states but the same heavy states.

\section{A CFT correlator at the orbifold point}
\label{sec:orb-cft}

At the orbifold point the D1-D5 CFT can be described in terms of elementary free fields. In this work we focus on the untwisted sector where we have $N$ groups of  bosons and fermions\footnote{We follow the conventions of~\cite{Galliani:2016cai}, which are based on~\cite{Avery:2010qw}.} labelled by an index $r=1,\ldots, N$ 
\begin{equation}
  \label{eq:elefields}
  \Big(X^{A \dot{A}}_{(r)} (\tau, \sigma)\;,~ \psi_{(r)}^{\alpha \dot{A}}(\tau+\sigma)\;,~ \tilde{\psi}_{(r)}^{\dot{\alpha} \dot{A}}(\tau-\sigma)\Big)~,
\end{equation}
where the indices $(\alpha\,\dot{\alpha})$ are in the fundamental representation of the $SU(2)_L \times SU(2)_R$ which are part of the R-symmetry group, while $A$ and $\dot{A}$ are in the fundamental representation of two other $SU(2)$'s that we call $SU(2)_B$ and $SU(2)_C$ respectively  for ``bonus'' and ``custodial'' ($SU(2)_B$ is an outer automorphism of the superalgebra). 

As standard in an orbifold description, we have to keep only states invariant under the orbifold group, so the operators involved in the correlators must be invariant under the $S_N$ transformations permuting the copies of ${\cal M}$. In the untwisted sector this is achieved simply by symmetries over the index $r$; for instance we will consider the following operators in the NS-NS sector
\begin{equation}
  \label{eq:OL}
  O^{\alpha \dot\beta} =  \sum_{r=1}^N O_{(r)}^{\alpha \dot\beta} = \sum_{r=1}^N \frac{-i}{\sqrt{2 N}} \psi_{(r)}^{\alpha \dot{A}}\, \epsilon_{\dot{A} \dot{B}}\, \tilde{\psi}_{(r)}^{\dot{\beta} \dot{B}}~,~~~ J^3 = - \frac{1}{2} \sum_{r=1}^N \psi_{(r)}^{+ \dot{A}}\, \epsilon_{\dot{A} \dot{B}}\, {\psi}_{(r)}^{- \dot{B}}~.
\end{equation}
These operators are protected also away from the orbifold point since they are part chiral-primary multiplets, i.e. the highest weight state conformal dimension is equal to the R-symmetry spin $j$ (defined as the eigenvalue under $J^3$): $h=j=\bar{h}=\bar{j}=1/2$ for $ O^{+ +}$ and $h=j=1$, $\bar{h}=\bar{j}=0$ for $J^+$. These operators are light since their conformal dimension $\Delta=h+\bar{h}$ remains fixed when the central charge $c$ is scaled to infinity. On the contrary the R-R ground states are heavy since they have $h=\bar{h}=c/24$. 

We are interested in 4-point correlators with two light NS-NS operators and two R-R ground states\footnote{The plane and the cylinder coordinates are related by the change of variables $z=e^{\tau_E + i \sigma}$, $\bar{z}=e^{\tau_E - i \sigma}$, with $\tau = -i \tau_E$.}
\begin{equation}
  \label{eq:corrHHLL}
  \langle O_H(z_1,\bar{z}_1) \bar{O}_H(z_2,\bar{z}_2)
 O_L(z_3,\bar{z}_3) \bar{O}_L(z_4,\bar{z}_4) \rangle = \frac{1}{z_{12}^{2 h_H} \bar{z}_{12}^{2 \bar{h}_H}} \frac{1}{z_{34}^{2 h_L} \bar{z}_{34}^{2 \bar{h}_L}} {\cal G}(z,\bar{z})~,
\end{equation}
where $z_{jk} = z_j - z_k$ and ${\cal G}$ is a function of the projective-invariant ratio
\begin{equation}
  \label{eq:pirat}
  z = \frac{z_{14} z_{23}}{z_{13} z_{24}}~,~~~
  \bar{z} = \frac{\bar{z}_{14} \bar{z}_{23}}{\bar{z}_{13} \bar{z}_{24}}~.
\end{equation}
We can characterise the R-R insertions in terms of states (by sending $z_2 \to \infty$ and $z_1 \to 0$). A simple example of a correlator of the type~\eqref{eq:corrHHLL} is obtained by taking as the heavy operator $O_H$ the R-R state that has maximum value of the spin $j=N/2$. This state is related to the $SL(2,C)$-invariant vacuum by a spectral flow transformation and, in the orbifold language, correspond to the product of the R-R ground state $|++ \rangle_{(r)}$ of spin $(j=1/2,\bar{j}=1/2)$ in each copy of the CFT. If we choose the heavy and light operator as follows
\begin{equation}
  \label{eq:LMex}
  O_L = O^{++}~,~~~ \lim_{z\to 0}O_H |0\rangle = \prod_{r=1}^N |++ \rangle_{(r)} \equiv |++\rangle^N~,
\end{equation}
the correlator~\eqref{eq:corrHHLL} takes the following form ${\cal G}(z,\bar{z})=\frac{1}{|z|}$. In~\cite{Galliani:2016cai}, this correlator was analysed both within the CFT and the dual supergravity description, and it was shown that the two results agree. This non-renormalization property can be understood by decomposing the result in the channel where the two light operators approach each other and by showing that the correlator is saturated by considering the $U(1)$-affine descendants of $(J^3,\tilde{J}^3)$. The same result holds for a more general class of correlators~\cite{Galliani:2016cai}, where the heavy states are different from the ones in~\eqref{eq:LMex} but share a key property: as in~\eqref{eq:LMex} they are constructed by multiplying the {\em same} building block which acts on different copies of the CFT. The building blocks considered in~\cite{Galliani:2016cai} live in the $k^{\rm th}$ twisted sector of the orbifold CFT and can also carry a (holomorphic) momentum obtained by taking a spectral flow of level $s$ (in the holomorphic sector): again the correlators~\eqref{eq:corrHHLL} constructed with these heavy operators are protected and the orbifold CFT results match the corresponding supergravity expressions.

In this paper we focus on a different class of heavy operators whose basic constituents are not all identical. An interesting operator of this type can be constructed by taking $N-p$ copies in the state $|++\rangle_{(r)}$ as in~\eqref{eq:LMex} and the remaining copies in a R-R ground state $|00\rangle_{(r)}$ of spin $(j,\bar{j})=(0,0)$
\begin{equation}
  \label{eq:00}
  |00\rangle_{(r)} = \lim_{z,\bar{z}\to 0} 
O^{--}_{(r)}(z,\bar{z})\, |++\rangle_{(r)}~.
\end{equation}
In order to obtain a heavy state that has semiclassical dual description as a smooth geometry, one needs to take a linear combination of such states with a different number $p$ of $|00\rangle_{(r)}$ constituents~\cite{Skenderis:2006ah}
\begin{equation}
  \label{eq:SKs}
  |s_B\rangle = \frac{1}{N^{\frac{N}{2}}} \sum_{p=0}^N A^{N-p}\, |++\rangle^{N-p}\; B^{p}\, |00\rangle^p~,~~~{\rm with}~~~  |A|^2 + |B|^2 = N~,
\end{equation}
where $A$ and $B$ are complex parameters. Here we follow the conventions of~\cite{Giusto:2015dfa}: we understand a full symmetrization between the $N$ copies in~\eqref{eq:SKs} and the norm of the ket $|++\rangle^{N-p}\, |00\rangle^p$ is equal to $\binom{N}{p}$, i.e. the number of distinct permutations of the constituents. When $A$ and $B$ are of order $\sqrt{N}$, then the sum over $p$ is peaked, in the large $N$ limit, around $p\sim |B|^2$ and this semiclassical state is dual to a smooth $1/4$-BPS geometry. In this paper we study in detail the correlator~\eqref{eq:corrHHLL} where the light operator is as in~\eqref{eq:LMex}, while the heavy one creates the ket~\eqref{eq:skex}; in summary we choose
\begin{equation}
  \label{eq:skex}
  O_L = O^{++}~,~~~ \lim_{z\to 0}O_H |0\rangle = |s_B\rangle ~.
\end{equation}
It is straightforward to calculate this correlator at the free orbifold point and, with the choice~\eqref{eq:skex}, we obtain~\eqref{eq:corrHHLL} with
\begin{equation}
  \label{eq:Gsk}
{\cal G}(z,\bar{z}) = 
\frac{1}{|z|} + \frac{|B|^2}{2N}\, \frac{|z|^2+|1-z|^2-1}{|z|} 
+ \frac{|A|^2 |B|^2}{N} \left(1-\frac{1}{N}\right) \frac{|1-z|^2}{|z|}~.
\end{equation}
Notice that the last term scales, in the large $N$ limit, as ${\cal O}(N)$, while the first two terms are of order ${\cal O}(N^0)$. In the free CFT calculation they have two different combinatoric origins: the leading term in $N$ is due to the contributions from the terms where the light operators $O_L$ and $\bar{O}_L$ are non trivial in different copies, while the remaining terms are due to the ``diagonal'' contribution where both $O_L$ and $\bar{O}_L$ act on the same copy.

\section{Gravity}
In this section we describe the holographic computation of the correlator (\ref{eq:Gsk}). We first introduce the general background geometries and the linearized equations satisfied by the perturbation describing the light operator $O_L$, and then specialise to the geometry dual to the state $|s_B\rangle$ in (\ref{eq:SKs}).
\label{sec:gravity}

\subsection{The D1-D5 background}
\label{sec:d1-d5back}

We specialize to D1-D5 states that are invariant under rotations of the compact 4D manifold $\mathcal{M}$. Their dual geometries can be described in the 6D theory\footnote{A nice review of this theory can be found in appendix B of \cite{Roy:2016zzv}.}  reduced on $\mathcal{M}$ and have the general form \cite{Lunin:2001jy,Lunin:2002iz,Kanitscheider:2007wq,Giusto:2013rxa}:
\begin{subequations}
\label{eq:backgroundgeneral}
\begin{align}
&ds^2_{6}=-\frac{2}{\sqrt{\mathcal{P}}}(dv+\beta)\left(du+\omega\right)+\sqrt{\mathcal{P}}\,ds^2_4\,,\quad \mathcal{P}=Z_1 Z_2 - Z_4^2\,,\\
&e^{2\phi_1} = \frac{Z_1^2}{\mathcal{P}}\,,\quad e^{2\phi_2} = \frac{Z_2^2}{\mathcal{P}}\,,\quad \chi_1=\frac{Z_4}{Z_1}\,,\quad \chi_2=\frac{Z_4}{Z_2}\,,\\
&B=-\frac{Z_4}{\mathcal{P}}(du+\omega)\wedge (dv+\beta) + \delta_2\,,\quad C=-\frac{Z_2}{\mathcal{P}}(du+\omega)\wedge (dv+\beta) + \gamma_2\,.\label{eq:B&C}
\end{align}
\end{subequations}
$ds^2_6$ is the 6D Einstein metric, $\phi_1$ is the dilaton, $e^{\phi_1-\phi_2}$ is the volume of $\mathcal{M}$, $\chi_1$ is the RR 0-form, $\chi_2$ is the component along $\mathcal{M}$ of the RR 4-form, $B$ and $C$ are the NSNS and RR 2-forms. The 6D space is split in the light-cone coordinates
\begin{equation}
u = \frac{t-y}{\sqrt{2}}\,,\quad v = \frac{t+y}{\sqrt{2}}\,,
\end{equation}
with $t$ and $y$ the time and the $S^1$ coordinate, and the non-compact Euclidean 4D space $\mathrm{R}^4$, on which we define the flat metric $ds^2_4$. The different 2-charge D1-D5 microstates are characterized by the three scalars $Z_1$, $Z_2$, $Z_4$, the two 1-forms on $\mathrm{R}^4$ $\beta$, $\omega$ and the two 2-forms on $\mathrm{R}^4$ $\gamma_2$, $\delta_2$. All these are functions of the $\mathrm{R}^4$ coordinates but are independent of $u$ and $v$ and satisfy the differential relations
\begin{equation}
d\beta = *_4 d\beta\,,\quad d\omega = -*_4 d\omega\,,\quad d*_4 dZ_1=0\,,\quad *_4 dZ_2 = d\gamma_2\,,\quad *_4 dZ_4 = d\delta_2\,,
\end{equation}
with $*_4$ the Hodge dual with respect to $ds^2_4$. The expressions for the NSNS and RR 3-form field strengths will also be useful:
\begin{subequations}
\begin{align}
H &\equiv d B = \frac{Z_4 d(Z_1 Z_2) - (Z_1 Z_2 + Z_4^2) dZ_4}{\mathcal{P}^2}\wedge d\hat u\wedge d\hat v -\frac{Z_4}{\mathcal{P}}\left[ d\omega \wedge d\hat v - d\beta\wedge d\hat u\right]+*_4 dZ_4\,,\\
F&\equiv dC-\chi_1 H \nonumber \\
&=\left[\frac{Z_2}{Z_1} dZ_1 - \frac{Z_4}{Z_1} dZ_4\right]\wedge \frac{d\hat u \wedge d\hat v}{\mathcal{P}}-\frac{1}{Z_1}\left[ d\omega \wedge d\hat v - d\beta\wedge d\hat u\right]+*_4 d Z_2 - \frac{Z_4}{Z_1}*_4 dZ_4\,,
\end{align}
\end{subequations}
where we have defined
\begin{equation}
d\hat u \equiv du+\omega\,,\quad d\hat v\equiv dv+\beta\,.
\end{equation}
We note the following identities, that will be used in the next section: 
\begin{equation}
\label{eq:starH}
e^{-(\phi_1+\phi_2)}(H-*H) = \chi_2 (F-*F)\,,\quad e^{\phi_1-\phi_2} * F - \chi_2 H = - d \tilde C\,,
\end{equation}
where we denote by $*$ the Hodge-dual with respect to $ds^2_6$ with a 6D orientation such that $\epsilon_{uv1234}=+1$ and
\begin{equation}
\tilde C = -\frac{Z_1}{\mathcal{P}}(du+\omega)\wedge (dv+\beta) + \gamma_1\,,\quad *_4 d Z_1 = d \gamma_1\,.\label{eq:tildeC}
\end{equation}

\subsection{The perturbation}
\label{sec:perturbation}
The 6D fields dual to the (anti)-chiral primary operators $O_L$, ${\bar O}_L$ are a scalar $\mathrm{w}$ and a closed 3-form $h$. The linearized perturbation equations around the AdS$_3\times S^3$ background were derived in \cite{Deger:1998nm}. The AdS$_3\times S^3$ geometry is a special case of (\ref{eq:backgroundgeneral}) with $Z_4=\delta_2=0$, constant $\phi_1=-\phi_2$ and anti-self-dual $F$. Around such a background the perturbation equations for $(\mathrm{w},h)$ are
\begin{equation}\label{eq:pertzero}
h - *h = 2\, \mathrm{w}  F\,,\quad e^{2\phi_2}\,d * d \mathrm{w} = h\wedge F\,.
\end{equation}
One can identify $\mathrm{w}$ with the fluctuation of $\chi_2=e^{\phi_1-\phi_2}\chi_1$ and $h$ with the fluctuation of $H$.
 
We need to generalize the perturbation equations to a generic D1-D5 background. So we are looking for deformations of the solution (\ref{eq:backgroundgeneral}) controlled by a scalar $\mathrm{w}$ and a 3-form $h$ that satisfy the equations of motion at linear order, but generically break supersymmetry. It is immediate to see that in the presence of non-vanishing background values for $\chi_1$, $\chi_2$ and $H$, perturbing $\chi_1$, $\chi_2$ and $H$ induces at first order a perturbation of all other fields, so the task of constructing a consistent deformation is considerably more involved in this more general setting. Even if the perturbed solution does not need to be supersymmetric, we can use the supersymmetric solution (\ref{eq:backgroundgeneral}) as a guide to understand which fields will be excited by the perturbation. In particular we can consider the effect of varying at first order $Z_4$ by $\delta Z_4 \equiv \mathrm{w}\,Z_2$. This motivates the following ansatz for the perturbation:
\begin{equation}
\delta \chi_1 = e^{\phi_2-\phi_1} \mathrm{w} \,,\quad \delta \chi_2 = \mathrm{w}\,,\quad \delta e^{2\phi_2}= 2 e^{4\phi_2}\chi_2 \mathrm{w}\,,\quad \delta e^{2\phi_1}= 2 e^{2(\phi_1+\phi_2)}\chi_2 \mathrm{w}\,.
\end{equation}
Given the form of $B$ in (\ref{eq:B&C}), it is also natural to define $h$ as
\begin{equation}
\delta H \equiv h + d y \,,\quad y \equiv -2 \frac{Z_2 Z_4^2}{\mathcal{P}^2}\,\mathrm{w}\, d\hat u \wedge d\hat v\,.
\end{equation}
The Bianchi identity $dH=0$ implies $dh=0$. The Bianchi identity for $F$:
\begin{equation}
d (F+ \chi_1 H)=0
\end{equation}
implies the form of the $F$-variation:
\begin{equation}
\delta F = -e^{\phi_2-\phi_1} \mathrm{w} H - \chi_1 h -\chi_1 d y + dx\,.
\end{equation}
The 2-form $x$ is not fixed by the Bianchi identity, but the supersymmetric solution suggests the ansatz
\begin{equation}
x = -2 \frac{Z_2^2 Z_4}{\mathcal{P}^2}\,\mathrm{w} \,d\hat u \wedge d\hat v = \chi_2^{-1} y\,,
\end{equation}
which is what one would obtain by varying $Z_4$ in the expression for $C$ in (\ref{eq:B&C}). Since $Z_4$ also appears in the 6D Einstein metric, $ds^2_6$ should also fluctuate. Instead of guessing the full form of the metric perturbation, we will determine its effects on the 6D Hodge star by consistency with the equations of motion.  The Maxwell's equation for $F$
\begin{equation}
d(e^{\phi_1-\phi_2} *F - \chi_2 H)=0
\end{equation}
implies
\begin{equation}\label{eq:starF}
\delta (* F) = e^{\phi_2-\phi_1} \mathrm{w} H + \chi_1 h +\chi_1 d y - e^{\phi_2-\phi_1}d\tilde x\,;
\end{equation}
the 2-form $\tilde x$, which represents the variation of the dual potential $\tilde C$, can be inferred from (\ref{eq:tildeC}):
\begin{equation}
\tilde x =  -2 \frac{Z_1 Z_2 Z_4}{\mathcal{P}^2}\,\mathrm{w} d\hat u \wedge d\hat v =e^{\phi_1-\phi_2} x\,.
\end{equation}
If one assumes that (\ref{eq:starH}) is preserved by the perturbation, one deduces 
\begin{equation}
\begin{aligned}
e^{-(\phi_1+\phi_2)}\delta(* H) &=(2- e^{-(\phi_1+\phi_2)})h-\mathrm{w}(1+2 e^{2\phi_2}\chi_2^2)(F-*F) \\&+ 2 \chi_2 e^{\phi_2-\phi_1}\mathrm{w} H + (2- e^{-(\phi_1+\phi_2)}) dy + \chi_2 (dx+d\tilde x)\,.
\end{aligned}
\end{equation}
Finally, we need to know how the Hodge star acting on 1-form on $\mathrm{R}^4$ is deformed: since in the supersymmetric ansatz one does not get any factor of $Z_4$ when the star acts on 1-forms with legs only along the spatial directions, we assume that $\delta (* \omega_1) = *\, \delta \omega_1$ for any 1-form $\omega_1$ on $\mathrm{R}^4$.

We can now apply these deformation rules on the remaining equations of motion and require that they are preserved at first order in the deformation. If one looks at the variation of the equations for the RR scalars:
\begin{equation}
d(e^{2\phi_1}* d\chi_1)-e^{\phi_1-\phi_2}*F \wedge H=0\,,\quad d(e^{2\phi_2}* d\chi_2)+F \wedge H=0\,,
\end{equation}
one finds that $\mathrm{w}$ and $h$ must satisfy the differential constraints 
\begin{equation}\label{eq:pert}
e^{-(\phi_1+\phi_2)}(h-*h) = \mathrm{w} (F-*F)\,,\quad d(e^{2\phi_2} * d\mathrm{w}) + dC \wedge h=0\,. 
\end{equation}
These are the perturbation equations that generalize (\ref{eq:pertzero}) around a general D1-D5 background.

As a further consistency check, one can verify that the identity 
\begin{equation}
H\wedge *F + * H \wedge F=0\,,
\end{equation}
is preserved by our deformation rules: this checks that the deformation ansatz for the Hodge start operation, that we have derived somewhat indirectly (see eqs. (\ref{eq:starF}) and (\ref{eq:starH})), is actually consistent.
 
 \subsection{A particular D1-D5 microstate}
 The 4-point function of the light operators $O_L$, $\bar O_L$ and the heavy operators $O_H$, $\bar O_H$ is found by solving the perturbation equations (\ref{eq:pert}) in the background sourced by the heavy operators. When $O_H$ is the operator in (\ref{eq:skex}), the metric functions that describe the associated geometry are collected in Appendix \ref{sec:d1d5metric}. The geometry depends on two parameters $a$ and $b$, that are related to the CFT parameters $A$ and $B$ via \cite{Giusto:2015dfa}
 \begin{equation}
\label{eq:AB}
 A = R\, \sqrt{\frac{N}{Q_1 Q_5}}\,a\,,\quad  B = R\, \sqrt{\frac{N}{2\,Q_1 Q_5}}\,b\,.
 \end{equation}
 Here $Q_1$ and $Q_5$ are the supergravity D1 and D5 charges, given in terms of the numbers $n_1$, $n_5$ of D1, D5 branes by the usual relations
 \begin{equation}
Q_1 = \frac{(2\pi)^4 \,n_1\,g_s\,\alpha'^3}{V_4}\,,\quad Q_5 = n_5\,g_s\,\alpha'\,,
\end{equation}
with $g_s$ the string coupling and $V_4$ the volume of $\mathcal{M}$; $R$ is the radius of the $S^1$ on which the CFT lives. The constraint $|A|^2+|B|^2=N$ translates into
 \begin{equation}\label{eq:abconstraint}
 \frac{Q_1 Q_5}{R^2}=a^2 + \frac{b^2}{2}\,. 
 \end{equation}
 When $b=0$ the geometry is just AdS$_3\times S^3$:
 \begin{subequations}\label{eq:AdS3xS3}
\begin{align} 
&ds^2_6=\sqrt{Q_1 Q_5}\,(ds^2_{AdS_3}+ds^2_{S^3})\,,\\
&ds^2_{AdS_3}=\frac{d r^2}{a_0^2+r^2}-\frac{a_0^2+r^2}{Q_1Q_5}dt^2+\frac{r^2}{Q_1Q_5}dy^2\,,\quad a_0 \equiv \frac{\sqrt{Q_1 Q_5}}{R}\,,\label{eq:AdS3} \\
&ds^2_{S^3}=d \theta^2+\sin^2\theta\, d\hat{\phi}^2+\cos^2\theta\, d\hat{\psi}^2\,,\quad \hat\phi \equiv \phi -\frac{t}{R}\,\,,\,\,\hat \psi \equiv \psi- \frac{y}{R}\,, \label{eq:S3}\\
&C=C_{0}=-\frac{r^2}{Q_1}\,dt\wedge dy - Q_5\,\cos^2\theta\,d\hat \phi\wedge d\hat \psi\,,\quad e^{2\phi_1}= e^{-2\phi_2}= \frac{Q_1}{Q_5}\,,
\end{align}
\end{subequations}
with all other fields vanishing. When $b$ does not vanish the geometry cannot be factorized in an AdS$_3$ and an $S^3$ part, and solving (\ref{eq:pert}) in such a background seems a daunting task. Since the CFT result (\ref{eq:Gsk}) shows that the first non-trivial contribution to the correlator appears at order $b^2$, we set up the gravity computation through a perturbative expansion in $b$, and we concentrate on the terms of order $b^2$. We thus expand both the background and the perturbation up to order $b^2$, and we organize the computation using the AdS$_3$ plus $S^3$ split of the $b=0$ background.  We perform the expansion in $b$ keeping the D1, D5 charges $Q_1$, $Q_5$ and the $S^1$ radius $R$ fixed; from the CFT point of view this is the most natural expansion, as it changes neither the central charge $c=6\, n_1 n_5$, nor the CFT spatial volume. Note that, according to the constraint (\ref{eq:abconstraint}), the parameter $a$ has to be varied when expanding in $b$.

The background solution expanded up to order $b^2$ is given in Appendix \ref{sec:d1d5metric}. At linear order in $b$ the deformation of the background around AdS$_3\times S^3$ is controlled by the scalars $\chi_1$, $\chi_2$ and by the 2-form $B$: with the identifications $\mathrm{w}=\chi_2=Q_1/Q_5 \chi_1$ and $h= d B$, this is the deformation described in section \ref{sec:perturbation}, and it satisfies eq. (\ref{eq:pertzero}). The expressions for $\chi_2$ and $B$ in (\ref{eq:chis}) and (\ref{eq:B}) are indeed linear combinations of two solutions of the form of eq. (4.15) of \cite{Galliani:2016cai}. The two solutions are determined by the AdS$_3$ scalars $B_{\pm}$ and by the $S^3$ scalar spherical harmonics $Y^{\pm\pm}$, which satisfy 
\begin{equation}
\Box_{AdS_3} B_\pm + B_\pm =0\,,\quad \Box_{S^3} Y^{\pm\pm}+3\, Y^{\pm\pm}=0\,,
\end{equation}
with
\begin{equation}
\Box_g \equiv \frac{1}{\sqrt{|g|}} \partial_\mu \sqrt{|g|} \,g^{\mu\nu} \partial_\nu\,.
\end{equation}
Their explicit form is
\begin{equation}\label{eq:Bpm}
B_{\pm}=  \frac{a_0}{\sqrt{r^2+a_0^2}}\,e^{\pm i \,t/R}\,,\quad Y^{\pm\pm} = \sin\theta \,e^{\pm i \hat \phi}\,.
\end{equation} 
The $\mathcal{O}(b^2)$ terms in the metric, in the RR 2-form and in the scalars $\phi_1$, $\phi_2$ represent the backreaction, at second order in $b$, of this linear perturbation. They are encoded in the AdS$_3$ metric fluctuation $\delta g_{\mu\nu}$ and in the vectors $A^\phi$ and $A^\psi$ 
\begin{equation}
\delta g_{\mu\nu} \,dx^\mu dx^\nu = \frac{dt^2}{Q_1 Q_5}\,, \quad A^\phi=\frac{R}{2\,Q_1 Q_5}\,dt \,,\quad A^\psi=\frac{R}{2\,Q_1 Q_5}\,\frac{r^2}{r^2+a_0^2}\,dy\,,
\end{equation}
which are linked to the first order perturbation fields $B_{\pm}$ by covariant differential identities:
\begin{subequations}
\begin{align}\label{eq:bkgcoefficients}
&\delta g^\mu_\mu = - a_0^{-2}\,B_+ B_-\,,\quad \nabla^\mu\! \left(\!\delta g_{\mu\nu}-\frac{1}{2} g_{\mu\nu} \,\delta g^\rho_\rho \right)=0\,,\\
&(\Box_{AdS_3}+2)\, \delta g_{\mu\nu}= - a_0^{-2}\,(\partial_\mu B_+ \partial_\nu B_- + \partial_\nu B_+ \partial_\mu B_-)\,,\\
&d A^\phi=0\,,\quad d A^\psi = -\frac{i}{2\,a_0^2} \,(B_- *_{AdS_3} d B_+ -B_+ *_{AdS_3} d B_-)\,,
\end{align}
\end{subequations}
where to raise and lower indices and to define covariant derivatives one uses the unperturbed AdS$_3$ metric $ds^2_{AdS_3}$. The constant gauge field $A^\phi$ and the constant part of $A^\psi$ could be re-absorbed by a redefinition of the $\phi$ and $\psi$ coordinates, and hence are not determined by the equations of motion. These diffeomorphisms, however, do not vanish at the AdS boundary, and have a physical effect on the geometry: we will see that the result of the correlator, indeed, depends on $A^\phi$.

\subsection{Calculation of the 4-point function}
\label{sec:4pointcomp}

As we see from the CFT result (\ref{eq:Gsk}), the correlator comprises a term of order $N$, which dominates the large $N$ expansion, and a subleading term of order $N^0$:
\begin{equation}
\begin{aligned}
&\langle O_H(t=-\infty)  {\bar O}_H(t=\infty) O_L(0,0) \bar O_L(t,y) \rangle\equiv \frac{e^{-i \frac{t}{R}}}{N}\,|\langle O_H(t=-\infty)  {\bar O}_H(t=\infty) O_L(0,0)\rangle|^2\\
&\qquad\qquad+\langle O_H(t=-\infty)  {\bar O}_H(t=\infty) O_L(0,0) \bar O_L(t,y) \rangle^{(0)}\,.
\end{aligned}
\end{equation}
 The leading term, which represents the disconnected part of the correlator, is proportional to the modulus square of the 3-point function \cite{Kanitscheider:2007wq,Giusto:2015dfa}
\begin{equation}
\langle O_H(t=-\infty)  {\bar O}_H(t=\infty) O_L(0,0)\rangle = B \,\bar A\,,
\end{equation}
a protected quantity which can be computed both in supergravity and at the free orbifold point. The order $N^0$ term, which is denoted with the subscript $(0)$, is the sum of the subleading part of the disconnected correlator and of the connected correlator, and is, in general, a non-protected quantity. To compute this term on the gravity side, one needs to solve the linearized equations (\ref{eq:pert}) for $(\mathrm{w},h)$ in the background (\ref{eq:bkg2ndorder}), with the following boundary condition for $\mathrm{w}$ at large $r$: 
\begin{equation}
\label{eq:pert++}
\mathrm{w}\approx \delta(t,y)\,\frac{\log r}{r}\,Y^{++}(\theta,\phi) + \frac{b(t,y)}{r}\,Y^{++}(\theta,\phi) + \ldots\,;
\end{equation}
the term proportional to $\log r/r$ is the source for $O_L$ localized at the point $t=y=0$ on the boundary, the term proportional to $1/r$ is the normalizable term proportional to the vev of ${\bar O}_L$, and the dots represent terms proportional to spherical harmonics other than $Y^{++}$, which do not contribute to the correlator of interest here. We further require $\mathrm{w}$ and $h$ to be regular in the interior of the space, and this determines uniquely the normalizable term. The vev of ${\bar O}_L$ in the presence of a source for $O_L$ gives the 2-point function in the geometry sourced by $O_H$, and hence the correlator at order $N^0$ is given by
\begin{equation}
\langle O_H(t=-\infty)  {\bar O}_H(t=\infty) O_L(0,0) \bar O_L(t,y) \rangle^{(0)} = b(t,y)\,,
\end{equation}
up to a proportionality factor of which we do not keep track here. The correlator above is computed on the cylinder with coordinates $t, y$. We can transform from the cylinder to the Minkowski plane with the usual transformation
\begin{equation}
z = e^{i \frac{t+y}{R}}\,,\quad {\bar z} = e^{i \frac{t-y}{R}}\,,\quad {\bar O}_L(z,\bar z) = (z \bar z)^{-1/2}\,{\bar O}_L(t,y)\,,
\end{equation}
and we can further analytically continue to the Euclidean complex plane by sending $t\to - i t_e$, with $t_e$ the Euclidean time. Then the $\mathcal{O}(N^0)$ correlator on the plane is
\begin{equation}\label{eq:cylinder2plane}
\langle O_H(0) \bar O_H(\infty) O_L(1,1)  \bar O_L(z,\bar z) \rangle^{(0)} = \frac{1}{|1-z|^2} \,\mathcal{G}^{(0)}(z, \bar z) = (z \bar z)^{-1/2}\,b(z,\bar z)\,,
\end{equation}
where $\bar O_H(\infty)$, $O_H(0)$ denote the heavy operators evaluated at $z=\infty$ and $z=0$.

The $b$-expansion of the background induces a corresponding expansion for the perturbation $(\mathrm{w},h)$:
\begin{equation}
\mathrm{w} = \mathrm{w}_0 + b^2 \,\mathrm{w}_1 + \mathcal{O}(b^4)\,,\quad h = h_0 + b^2\,h_1 + \mathcal{O}(b^4)\,.
\end{equation}
The order zero term $(\mathrm{w}_0,h_0)$ is the solution of (\ref{eq:pert}) in the AdS$_3\times S^3$ background (\ref{eq:AdS3xS3}) and hence it admits the factorized form \cite{Galliani:2016cai}
\begin{equation}\label{eq:zeropert}
\mathrm{w}_0 = B_0(r,t,y)\,Y^{++}(\theta,\hat \phi)\,,\quad h_0 = Q_5\,d \,[Y^{++}(\theta,\hat \phi) *_{AdS_3} d B_0 - B_0 *_{S^3} d Y^{++}(\theta,\hat\phi)]\,,
\end{equation}
where $*_{AdS_3}$ and $*_{S^3}$ are the Hodge duals with respect to the AdS$_3$ and $S^3$ metrics (\ref{eq:AdS3}), (\ref{eq:S3}). The AdS$_3$ function $B_0(r,t,y)$ satisfies $\Box_{AdS_3} B_0 + B_0=0$, has the boundary behaviour $B_0 \approx \delta(t,y) \log r/r$ for large $r$ and is regular at any finite value of $r$: it is thus the usual bulk-to-boundary propagator for a field of dimension $\Delta=1$ in global AdS$_3$, with the boundary point set to $t'=y'=0$:
\begin{equation}\label{eq:B0}
B_0(r,t,y) = K_1^{\mathrm{Glob}}(r,t,y|t'=0,y'=0)=\frac{1}{2}\,\frac{a_0}{\sqrt{r^2+a_0^2} \,\cos (t/R) - r\,\cos (y/R)}\,.
\end{equation}
To extract the order $b^0$ contribution to the correlator, we have to transform from the coordinate $\hat \phi$ to $\phi$ (which is equivalent to spectrally flowing from the NSNS to the RR sector of the CFT): this is easily done by using
\begin{equation}\label{eq:sffactor}
Y^{++}(\theta,\hat \phi) = e^{-i \frac{t}{R}}\,Y^{++}(\theta,\phi)\,.
\end{equation}
Taking the large $r$ limit of (\ref{eq:B0}), including the $e^{-i \frac{t}{R}}$ factor coming from (\ref{eq:sffactor}), and switching to Euclidean plane coordinates, we find the $b=0$ value of the correlator
\begin{equation}
\mathcal{G}(z,\bar z)|_{b=0} = \frac{1}{|z|}\,.
\end{equation}
This result was already obtained in \cite{Galliani:2016cai}, and coincides with the $b=0$ term of the CFT correlator. 

The interesting new information is contained in the $\mathcal{O}(b^2)$ terms $(\mathrm{w}_1,h_1)$. The perturbation equations (\ref{eq:pert}) at order $b^2$ give:
\begin{subequations}\label{eq:pert1}
\begin{align}\label{eq:pert1a}
&h_1 - *_0 \,h_1 - 2\, \mathrm{w}_1\,d C_0 =  \mathcal{F}~, \\
\label{eq:pert1b}
&\mathcal{F} \equiv \mathrm{w}_0 \,(F_1 - *_0 F_1- *_1 d C_0)+*_1 h_0 - 2\, (e^{-(\phi_1+\phi_2)})_1\,\mathrm{w}_0 \,dC_0\,,
\end{align}
\end{subequations}
and
\begin{subequations} \label{eq:pert12}
\begin{align} \label{eq:pert12a}
  &\frac{Q_5}{Q_1}\, d*_0 d \mathrm{w}_1 -  h_1 \wedge dC_0 = \widehat{\mathcal{F}} \\ \label{eq:pert12b}
&\widehat{\mathcal{F}}\equiv h_0 \wedge dC_1 - \frac{Q_5}{Q_1}\, d *_1 d \mathrm{w}_0 - d [ (e^{2\phi_2})_1 *_0 d \mathrm{w}_0]\,.
\end{align}
\end{subequations}
The left-hand sides of~\eqref{eq:pert1a} and~\eqref{eq:pert12a} are obtained by keeping the second order terms of $(\mathrm{w}, h)$ and the zeroth order background ($*_0$ is the Hodge dual with respect to the AdS$_3\times S^3$ metric (\ref{eq:AdS3xS3})); vice versa the sources in~\eqref{eq:pert1b} and~\eqref{eq:pert12b} originate from the zeroth order perturbation (\ref{eq:zeropert}) and the second order corrections to the background~\eqref{eq:bkg2ndorder}, which we denote with the subscript 1. The sources $\mathcal{F}$ and $\widehat{\mathcal{F}}$ contain scalar and vector spherical harmonics of various orders; we need to keep only terms containing the scalar harmonic $Y^{++}$ (or its derivative), which are the only ones contributing to our correlator. Some identities on $S^3$ spherical harmonics useful to perform the projection on $Y^{++}$ are collect in Appendix \ref{sec:harmonics}. A laborious computation gives:
\begin{subequations}\label{eq:sources}
\begin{align}
&\mathcal{F}= Q_5 \,[f_0\,(\mathrm{vol}_{S^3} - \mathrm{vol}_{AdS_3})\,Y^{++}(\theta,\hat \phi) + f_1\wedge *_{S^3} d Y^{++}(\theta,\hat \phi) + *_{AdS_3} f_1 \wedge d Y^{++}(\theta,\hat \phi)]\,,\\
&f_0=\frac{1}{6 a_0^2} B_0 B_+ B_- - i \,\partial^\mu B_0\,A^\phi_\mu\,,\\
&f_1 = \frac{1}{9 a_0^2} B_0 B_- d B_+ -\frac{2}{9 a_0^2} B_0 B_+ d B_--\frac{1}{18 a_0^2} B_+ B_- d B_0 - i\,B_0\,A^\phi+ \frac{2\,i}{3}\,*_{AdS_3}(d B_0\wedge A^\psi) \nonumber\\
&\qquad - \delta g_{\mu\nu} \partial^\mu B_0 dx^\nu\,,\\
&\widehat{\mathcal{F}}= Q_5^2\,\hat f_0\,Y^{++}(\theta,\hat \phi)\,\mathrm{vol}_{AdS_3}\wedge \mathrm{vol}_{S^3}\,,\\
&\hat f_0=-\frac{1}{3 a_0^2}B_0 B_+ B_- + \frac{1}{a_0^2} B_-\, \partial_\mu B_+ \,\partial^\mu B_0 -\nabla_\mu \partial_\nu B_0 \,\delta g^{\mu\nu} - 4 i \,\partial^\mu B_0\,A^\phi_\mu\,,
\end{align}
\end{subequations}
where operations on the AdS$_3$ indices $\mu,\nu$ are performed using the unperturbed AdS$_3$ metric (\ref{eq:AdS3}) and\
\begin{equation}
\mathrm{vol}_{AdS^3}=\frac{r}{Q_1 Q_5}\,dr\wedge dt\wedge dy\,,\quad \mathrm{vol}_{S^3}= \sin\theta\,\cos\theta\,d\theta\wedge d\hat \phi\wedge d\hat \psi\,.
\end{equation}
A general ansatz for $(\mathrm{w}_1, h_1)$, that only includes  the scalar spherical harmonic $Y^{++}$, is
\begin{equation}
\mathrm{w}_1 = B_1\,Y^{++}(\theta,\hat \phi)\,,\quad h_1 = Q_5\, d [S_1 *_{S^3} d Y^{++}(\theta,\hat \phi)+*_{AdS_3} V_1\,Y^{++}(\theta,\hat \phi) ]\,,
\end{equation}
where $B_1$, $S_1$ are scalars and $V_1$ is a 1-form on AdS$_3$. Note that we could have added to $h_1$ a term $Q_5\,d [\tilde V_1 \wedge d Y^{++}]$ for some 1-form $\tilde V_1$, but this amounts to redefining $*_{AdS_3} V_1\to *_{AdS_3} V_1+ d \tilde V_1$. Eqs. (\ref{eq:pert1}) then reduce to
\begin{equation}
-3\,S_1-\nabla_\mu V^\mu_1 - 4\,B_1=f_0\,,\quad dS_1 + V_1 = f_1\,,\quad \Box B_1-3 B_1 + 2 \nabla_\mu V^\mu_1-6 S_1 = - \hat f_0\,.
\end{equation}
One can solve the middle equation for $V_1$ and substitute it in the remaining two equations. These become coupled differential equations for the two scalars $B_1$ and $S_1$, which can be decoupled by introducing the combinations
\begin{equation}
s=B_1-3 S_1\,,\quad t = B_1 + S_1\,.
\end{equation}
We then obtain the equations
\begin{equation}\label{eq:s&t}
\Box s+ s= - (\hat f_0+f_0+3 \nabla_\mu f^\mu_1)\equiv J_s\,,\quad \Box t - 15 \,t  =-\hat f_0+3 f_0+ \nabla_\mu f^\mu_1\equiv J_t \,.
\end{equation}
We see that $s$ is a field dual to an operator of dimension 1, while the operator dual to $t$ has dimension $5$ and is a super-descendant of the chiral primary $O_L$. To obtain the correlator we are interested in, we should then set $t=0$, which gives 
\begin{equation}
B_1(r,t,y) = \frac{s(r,t,y)}{4}=-\frac{i}{4}\,\int d^3\mathbf{r}'\, \sqrt{-g_{AdS_3}}\,G^{\mathrm{Glob}}_1(\mathbf{r}'|r,t,y)\,J_s(\mathbf{r}')\,,
\end{equation} 
where $\mathbf{r}'\equiv \{r',t',y'\}$ is a point in AdS$_3$ and $G^{\mathrm{Glob}}_1(\mathbf{r}'|r,t,y)$ is the bulk-to-bulk propagator for a scalar field of mass $m^2=-1$ in global AdS$_3$, normalized such that $(\Box+1) G^{\mathrm{Glob}}_1 = i/\sqrt{-g_{AdS_3}}\, \delta$. In the $r\to \infty$ limit $G^{\mathrm{Glob}}_1$ is related with the bulk-to-boundary propagator $K^{\mathrm{Glob}}_1(\mathbf{r}'|t,y)$ normalized as in (\ref{eq:B0}) by (see for example eq. (6.12) in \cite{DHoker:2002nbb})
\begin{equation}
G^{\mathrm{Glob}}_1(\mathbf{r}'|r,t,y) \to \frac{a_0}{2\pi\,r}\,K^{\mathrm{Glob}}_1(\mathbf{r}'|t,y)\,.
\end{equation}
After including the factor originating from the spectral flow relation (\ref{eq:sffactor}) and continuing to Euclidean signature ($t\to -i t_e$), one finds the order $b^2$ contribution to the $\mathcal{O}(N^0)$ correlator on the Euclidean cylinder:
\begin{equation}\label{eq:correlatorsemifinalcylinder}
\langle O_H(t_e\!=\!-\infty)  {\bar O}_H(t_e\!=\!\infty) O_L(0,0) \bar O_L(t_e,y) \rangle^{(0)}_{b^2} \!=- \frac{b^2 e^{-\frac{t_e}{R}}}{8\pi}\!\!\int \!\!d^3\mathbf{r}'_e\, \sqrt{\bar g}\,K_1^\mathrm{Glob}(\mathbf{r}'_e|t_e, y) J_s(\mathbf{r}'_e) ,
\end{equation}
with $\mathbf{r}'_e\equiv \{r',t'_e,y'\}$ and $\bar g$ the metric of Euclidean AdS$_3$. The source $J_s(\mathbf{r}'_e)$ follows from (\ref{eq:s&t}), (\ref{eq:sources}) and (\ref{eq:bkgcoefficients}):
\begin{equation}\label{eq:Js}
\begin{aligned}
J_s(\mathbf{r}'_e)=&-\frac{1}{3 a_0^2}\,B_0 B_+ B_-+\frac{1}{3 a_0^2}\,B_0\,\partial'_\mu B_+\,\partial'^\mu B_- - \frac{11}{3 a_0^2}\,B_-\,\partial'_\mu B_+\,\partial'^\mu B_0\\
&+\frac{1}{12}\,B_+\,\partial'_\mu B_-\,\partial'^\mu B_0+4 \delta g^{\mu\nu}\,\nabla'_\mu \partial'_\nu B_0 +8 i\,\partial'^\mu B_0\,A^\phi_\mu\\
&=-\frac{1}{3 a_0^2}\,B_0 B_+ B_-+\frac{1}{3 a_0^2}\,B_0\,\partial'_\mu B_+\,\partial'^\mu B_- - \frac{5}{3 a_0^2}\,B_-\,\partial'_\mu B_+\,\partial'^\mu B_0\\
&+\frac{7}{3 a_0^2}\,B_+\,\partial'_\mu B_-\,\partial'^\mu B_0-4\frac{a_0^2\,R^2}{(r'^2+a_0^2)^2}\,\partial^2_{t'_e} B_0 + 4\frac{R}{r'^2+a_0^2}\,\partial_{t'_e}B_0\,,
\end{aligned}
\end{equation}
with $\partial'_\mu$ the derivative with respect to $\mathbf{r}'_e$. To simplify the correlator it is useful to switch to Poincar\'e coordinates $\mathbf{w}\equiv \{w_0,w,\bar w\}$ via the coordinate transformation 
\begin{equation}\label{eq:glob2poinc}
w_0=\frac{a_0}{\sqrt{r'^2+a_0^2}}\,e^{\frac{t'_e}{R}}\,,\quad w=\frac{r'}{\sqrt{r'^2+a_0^2}}\,e^{\frac{t'_e+i y'}{R}}\,,\quad  {\bar w}=\frac{r'}{\sqrt{r'^2+a_0^2}}\,e^{\frac{t'_e-i y'}{R}}\,.
\end{equation}
The bulk-to-boundary propagator of global AdS$_3$ $K^{\mathrm{Glob}}_1(\mathbf{r}'_e|t_e,y)$ is related to the bulk-to-boundary propagator in Poincar\'e coordinates $K_1(\mathbf{w}|z, \bar z)$ as
\begin{equation}
\label{eq:g2p}
K^{\mathrm{Glob}}_\Delta(\mathbf{r}'_e|t_e,y) = |z|^\Delta\, K_\Delta(\mathbf{w}|z,\bar z)\quad \mathrm{with}\quad K_\Delta(\mathbf{w}|z, \bar z)= \frac{w^\Delta_0}{(w_0^2+|w-z|^2)^\Delta}\,;
\end{equation} 
the factor $|z|^\Delta$ in this relation is precisely the factor that appears in the transformation from the cylinder to the plane in (\ref{eq:cylinder2plane}). 
The propagator $B_0$ in~\eqref{eq:correlatorsemifinalcylinder} can be rewritten in Poincar\'e coordinates by using~\eqref{eq:g2p} with $\Delta=1$. Similarly, for $B_+$ and $B_-$, we use the identification for the boundary points $z=\infty$, $z=0$; for a general $\Delta$ the relations are
\begin{subequations}
  \label{eq:g2b0i}
\begin{align}
  B_+^\Delta(\mathbf{r}'_e) = & \lim_{z\to \infty}|z|^{2 \Delta} K_\Delta(\mathbf{w}|z,\bar z) = \left(\frac{a_0 e^{\frac{t'_e}{R}}}{\sqrt{r'^2+a_0^2}}\right)^\Delta \,,
\\
B_-^\Delta(\mathbf{r}'_e) = & ~K_\Delta(\mathbf{w}|0)=  \left(\frac{a_0 e^{-\frac{t'_e}{R}}}{\sqrt{r'^2+a_0^2}}\right)^\Delta\,.
\end{align}
\end{subequations}
With manipulations similar to the ones used to simplify Witten's diagrams \cite{Liu:1998ty,Freedman:1998bj, D'Hoker:1999pj, DHoker:1999mqo} (summarised in Appendix \ref{sec:red2dintegrals}), we can rewrite the correlator (\ref{eq:correlatorsemifinalcylinder}) in terms of the integrals
\begin{equation}\label{eq:Dhat}
\begin{aligned}
{\hat D}_{\Delta_1\Delta_2 \Delta_3\Delta_4}&\equiv \lim_{z_2\to\infty} |z_2|^{2\Delta_2}\,D_{\Delta_1\Delta_2\Delta_3\Delta_4}(z_1=0,z_2=\infty,z_3=1,z_4=z)\\
 &=\lim_{z_2\to\infty} |z_2|^{2\Delta_2} \int d^3 \mathbf{w}\,\sqrt{\bar g}\,K_{\Delta_1}(\mathbf{w}|0)\,K_{\Delta_2}(\mathbf{w}|z_2,{\bar z}_2)\,K_{\Delta_3}(\mathbf{w}|1)\,K_{\Delta_4}(\mathbf{w}|z,\bar z)\,.
\end{aligned}
\end{equation} 
By using~\eqref{eq:sumI} in~\eqref{eq:correlatorsemifinalcylinder}, we have from~\eqref{eq:cylinder2plane}
\begin{equation}\label{eq:finalcorrelator}
\begin{aligned}
\mathcal{G}^{(0)}(z,\bar z)|_{b^2} = \frac{b^2}{a_0^2}\,\frac{1}{|z|}\,\left[\frac{|z|^2}{\pi}\,\hat D_{2211}-\frac{1}{2}\right]\,.
\end{aligned}
\end{equation}
Including the $b=0$ and the $\mathcal{O}(N)$ terms, one finds the total correlator up to order $b^2$:
\begin{equation} \label{eq:finalcorrelatorab}
\mathcal{G}_\mathrm{grav}(z,\bar z) = \frac{1}{|z|}\,\left[1+\frac{b^2}{a_0^2}\,\left(\frac{|z|^2}{\pi}\,\hat D_{2211}-\frac{1}{2}+\frac{N}{2}\,|1-z|^2\right)\right]\,.
\end{equation}
Clearly the gravity result differs from the free correlator (\ref{eq:Gsk}), and so this correlator is not protected for $b\not=0$. Of course by using the same supergravity solutions it is possible to derive the correlator with a different choice of the light states while keeping the heavy states unchanged: in Appendix~\ref{sec:Opmap} we provide the result for the correlator with $O_L=O^{+-}$.

\section{Discussion and conclusions}
\label{sec:outl}

In this paper we calculated a HHLL correlator~\eqref{eq:corrHHLL} in the context of the AdS$_3$/CFT$_2$ duality. The heavy operators are RR ground states and so they have conformal weight $h=\bar{h}=c/24$, while the light operators are chiral primaries with quantum numbers $h=j=\bar{h}=\bar{j}=1/2$. In particular we focus on a class of RR ground states for which the dual bulk description is known in terms of an explicit, regular supergravity solution. On the CFT side, it is straightforward to calculate the correlator at a special point of the moduli space where the CFT reduces to a free orbifold and the result is given by~\eqref{eq:Gsk}. On the bulk side, the calculation is more challenging and the result~\eqref{eq:finalcorrelatorab} is obtained in the limit $b^2 \ll a^2$. Notice that the approximation on $b$ is performed after taking the large $N$ limit, so we are taking a double scaling limit where $N$ is large and $b$ is small but constant; in other words the parameter $B^2$~\eqref{eq:AB} on the CFT side is always of order $N$.

By comparing the two results mentioned above, it is clear that the correlator under investigation is not protected.  It would be interesting to study explicitly how this
correlator changes in conformal perturbation theory when the orbifold description is deformed with the blow-up twist two modes; this is a technically challenging calculation as it requires to add two new external states to previous analysis which focused on the 2-point function (see for instance~\cite{Pakman:2009mi,Carson:2016uwf}). On the bulk side, the gravity result has a more intricate structure and should display several features such as Landau singularities, anomalous dimensions and couplings of double trace operators that are absent in the free orbifold result. We leave the analysis of these points to a future work and here we wish to mention just a few consistency checks, which can be performed easily on the supergravity result. 

The exchange of the operators $O_L$ and $\bar O_L$ should act on the correlator as $\mathcal{G}(z,\bar z)\to \mathcal{G}(z^{-1},\bar z^{-1})$, as can be seen from the definitions (\ref{eq:corrHHLL}) and (\ref{eq:pirat}). On the gravity side, the correlator with exchanged $O_L$ and $\bar O_L$ is computed by replacing $Y^{++}\to Y^{--}$ in (\ref{eq:pert++}), which is equivalent to sending $\phi\to -\phi$. As the background is invariant if one reverses at the same time the signs of $\phi$ and $t$, this exchange produces a result for the correlator of the form of eq. (\ref{eq:sumI}) with $t\to -t$, $I_3$ and $I_4$ interchanged and $I_6\to -I_6$. Using the identities of Appendix \ref{sec:red2dintegrals} and the symmetry (\ref{eq:hatDsymmetry}) of the function $\hat D_{2211}$, one can verify that the final result indeed equals $\mathcal{G}_\mathrm{grav}(z^{-1},\bar z^{-1})$. 

Of course an important consistency check is provided by the study of the OPE degeneration, as done for the AdS$_5$ case~\cite{D'Hoker:1999jp,Arutyunov:2000ku}.  As usual, the expansion of the gravity correlator in the various channels contains logarithmic terms, which capture the anomalous dimensions of the exchanged multiparticle states. In the $z\to 1$ limit the two light operators are close and the leading multiparticle state is $:\bar{O}_L O_L:$. Since the single particle constituents are a chiral and an anti-chiral operator, this multiparticle state is expected to gain an anomalous dimension at subleading order in $1/N$, encoded in the coefficient of the term $|1-z|^2 \log|1-z|^2$ in the expansion of $\mathcal{G}$. By looking at the sign of the relevant logarithm, it is clear that the anomalous dimension of the state $:\bar{O}_L O_L:$ will be positive. This is a somewhat surprising result since, to the best of our knowledge, all the previous holographic computations, mostly performed in 4D CFT's, have produced negative anomalous dimensions~\cite{Fitzpatrick:2011hh,Alday:2017gde}. It is interesting to study the logarithms also in the $z\to 0$ limit: in this channel the multiparticle operators exchanged between $\bar O_L$ and $O_H$ are composed by constituents of the same chirality, unless they contain both holomorphic and anti-holomorphic derivatives. This implies the absence of terms of the form $z^n \log |z|^2$ and $\bar z^n \log |z|^2$, for any integer $n\ge 0$: one can verify that this expectation is fulfilled by our result. 

It is possible to extract some interesting information also from the non-logarithmic terms of the OPE expansions. For instance one can check that the contributions due to the exchange of the lowest order protected operators in the various OPE channels are equal at the orbifold and the gravity point: in the $z\to 1$ channel one can match the contributions from the identity and the R-currents $J^3$ and $\bar J^3$, and in the $z\to 0$ channel the contribution of the lowest order anti-chiral multiparticle operator exchanged between $\bar O_L$ and $O_H$. 

Beyond the leading order the analysis is more complicated. In the OPE channel where the two light operators are close, the gravity result contains holomorphic corrections that can not be explained by the exchange of the affine descendants of the identity (and of course the same holds in the antiholomorphic sector). This suggests that some primary operators, that at the orbifold point can be constructed by using affine descendants on different CFT copies, have a fix conformal dimension, with $\bar{h}=0$, at large $N$. It would be of course interesting to verify or disprove this conjecture, since it is relevant to determine the singularity structure of the large $c$ correlators and can shed some light on the problem of information loss in the dual gravitational description, see~\cite{Fitzpatrick:2016ive,Galliani:2016cai,Fitzpatrick:2016mjq}.

\vspace{7mm}
 \noindent {\large \textbf{Acknowledgements} }

 \vspace{5mm} 

We would like to thank A. Bernamonti, M. Buican, S. Ramgoolam, L. Rastelli, M. Shigemori for useful discussions and correspondence. S.G. and R.R. wish thank the Galileo Galilei Institute for Theoretical Physics (GGI) for the hospitality during the program ``New Developments in AdS3/CFT2 Holography''. This research is partially supported by STFC (Grant ST/L000415/1, {\it String theory, gauge theory \& duality}), by the Padova University Project CPDA144437 and by INFN.

\appendix

\section{Some details on the free correlator}
\label{sec:free-corr}

In this appendix we provide some detail on the derivation of~\eqref{eq:Gsk}. As mentioned at the end of section~\ref{sec:orb-cft}, there are two type of contributions: the ``diagonal'' terms where $O_L$ and $\bar{O}_L$ are non-trivial on the same copy and the ``off-diagonal''ones where the two light operators act on different copies. We start from the second type of contributions, which produces the last term of~\eqref{eq:Gsk}. In this case, the calculation has the structure of a product of two 3-point correlators and we can derive the contribution proportional to $|A|^2 |B|^2$ in~\eqref{eq:Gsk} (which we indicate by ${\cal G}_{off}$) by choosing $z_2 \to \infty$, $z_1 \to 0$, $z_3=1$ and $z_4=z$
\begin{equation}
  \label{eq:coAB}
  \frac{1}{|1-z|^2} {\cal G}_{off}(z,\bar{z}) = \sum_{r\not= s} \langle s_B| O^{++}_{(r)}(1,1) O^{--}_{(s)}(z,\bar{z}) |s_B \rangle~.
\end{equation}
By using~\eqref{eq:SKs} and the fact that the zero-mode of $O^{++}_{(r)}$ turns the state $|00\rangle_{(r)}$ into $|++\rangle_{(r)}$ and vice versa for $O^{--}_{(r)}$ we have
\begin{equation}
  \label{eq:a2}
  \sum_{r\not= s} O^{++}_{(r)}(1,1) O^{--}_{(s)}(z,\bar{z}) |s_B\rangle = \frac{1}{|z|} \sum_{p=0}^N \frac{A^{N-p} B^p}{N^{\frac{N}{2}}} \left[p (N-p)|++\rangle^{N-p} |00\rangle^p + \ldots \right]~,
\end{equation}
where the dots stand for terms that contain copies of the CFT that are not in the $|++\rangle$ or $|00\rangle$ state and that we can ignore as they do not give any contribution to~\eqref{eq:coAB}. The factors of $p$ and $(N-p)$ in~\eqref{eq:a2} follow from the action of $O^{++}$ and $O^{--}$ on the different $|00\rangle$ and $|++\rangle$ copies respectively. Then, since the norm of the $|++\rangle^{N-p}\, |00\rangle^p$ is $\binom{N}{p}$, we have
\begin{equation}
  \label{eq:a3}
\begin{aligned}
  {\cal G}_{off}(z,\bar{z}) = & ~ \frac{|1-z|^2}{|z|} \sum_{p=0}^N  p (N-p)\,\frac{|A^2|^{N-p} |B^2|^p}{N^N} \, \binom{N}{p} \\ = & ~ \frac{|1-z|^2}{|z|} \frac{N (N-1) |A|^2 |B|^2 (|A|^2+|B|^2)^{N-2}}{N^N}~,
\end{aligned}
\end{equation}
which, as anticipated, yields the last term of~\eqref{eq:Gsk} after using the normalisation condition~\eqref{eq:SKs} $|A|^2+|B|^2=N$. 

The diagonal contribution follows from the building blocks
\begin{align}
  \label{eq:bbpp}
 &  {}_{(r)} \langle ++ |  O^{++}_{(r)}(1,1) O^{--}_{(r)}(z,\bar{z})| ++ \rangle_{(r)} = \frac{1}{|1-z|^2} \frac{1}{|z|}~,
\\   \label{eq:bb00}
& {}_{(r)} \langle 00 |  O^{++}_{(r)}(1,1) O^{--}_{(r)}(z,\bar{z})| 00 \rangle_{(r)} =   \frac{1}{|1-z|^2} \frac{1}{2} \frac{1}{|z|} \left(1 + |z|^2 + |1-z|^2 \right)~.
\end{align}
These results can be derived explicitly by using the bosonization formulae as done in~\cite{Galliani:2016cai} or equivalently by using the RR mode expansion for the fermions. Alternatively, one can reconstruct the correlators from their behaviour as $z\to 0,1,\infty$. In both cases there should be a simple pole in $1-z$ and $1-\bar{z}$ due to the fusion of $O^{++}_{(r)}$ and $O^{--}_{(r)}$ on the identity. Then the zero-modes of $O^{--}_{(r)}$ act non-trivially both on $| 00 \rangle_{(r)}$ and $|++\rangle_{(r)}$, so again the two equations should have the same $z\to 0$ limit proportional to $1/|z|$. In the $z\to\infty$ limit there is a difference: the zero-modes of $O^{--}_{(r)}$ act non-trivially on ${}_{(r)} \langle 00 |$ and so in this limit~\eqref{eq:bb00} should be proportional to $1/|z|$. On the contrary, the first modes of $O^{--}_{(r)}$ acting non-trivially on ${}_{(r)} \langle ++|$ are at level one, so~\eqref{eq:bbpp} should go as $1/|z|^3$ as $z\to\infty$. These constraints determine uniquely the sphere correlators above. Then the diagonal contributions to~\eqref{eq:SKs} are obtained by counting how many times the building blocks above appear:
\begin{equation}
\begin{aligned}
  \label{eq:coA}
  {\cal G}_{diag}^{A}(z,\bar{z}) = & \frac{1}{|z|} \sum_{p=0}^\infty
 (N-p)\,\frac{|A^2|^{N-p} |B^2|^p}{N^N} \, \binom{N}{p} \\ = & ~ \frac{1}{|z|} \frac{N |A|^2 (|A|^2+|B|^2)^{N-1}}{N^N}~,
\end{aligned}
\end{equation}
which is the contribution following from ~\eqref{eq:bbpp} and
\begin{equation}
\begin{aligned}
  \label{eq:coB}
  {\cal G}_{diag}^{B}(z,\bar{z}) = & \frac{1+|z|^2+|1-z|^2}{2 |z|} \sum_{p=0}^\infty p\,\frac{|A^2|^{N-p} |B^2|^p}{N^N} \, \binom{N}{p} \\ = & ~ \frac{1+|z|^2+|1-z|^2}{2 |z|} \frac{N |B|^2 (|A|^2+|B|^2)^{N-1}}{N^N}~,
\end{aligned}
\end{equation}
which is obtained from~\eqref{eq:bb00}. Summing ${\cal G}^A_{diag}$, ${\cal G}^B_{diag}$ and ${\cal G}_{off}$, we obtain~\eqref{eq:Gsk}.

\section{A D1-D5 geometry}
\label{sec:d1d5metric}

The supergravity solution dual to the state (\ref{eq:SKs}) can be written in the form (\ref{eq:backgroundgeneral}) with (see for example \cite{Giusto:2013bda})
\begin{subequations}\label{skenderistwocharge}
\begin{align}
d s^2_4 &= (r^2+a^2 \cos^2\theta)\Bigl(\frac{d r^2}{r^2+a^2}+ d\theta^2\Bigr)+(r^2+a^2)\sin^2\theta\,d\phi^2+r^2 \cos^2\theta\,d\psi^2\,,\label{ds4flat2}\\
\beta &=  \frac{R\,a^2}{\sqrt{2}\,(r^2+a^2 \cos^2\theta)}\,(\sin^2\theta\, d\phi - \cos^2\theta\,d\psi)\,,\\
Z_1 &= 1+\frac{R^2}{Q_5} \frac{a^2+\frac{b^2}{2}}{r^2+a^2 \cos^2\theta}+\frac{R^2\, a^2\, b^2}{2\,Q_5}\,\frac{\cos2\phi\,\sin^2\theta}{(r^2+a^2 \cos^2\theta)(r^2+a^2)}\,,\\
Z_2 &=  1+\frac{Q_5}{r^2+a^2 \cos^2\theta}\,,\quad \gamma_2 = -Q_5\,\frac{(r^2+a^2)\,\cos^2\theta}{r^2+a^2\cos^2\theta}\,d\phi\wedge d\psi\,,\\
Z_4 &=  R\, a\, b\,\frac{\cos\phi\,\sin\theta}{\sqrt{r^2+a^2}\,(r^2+a^2 \cos^2\theta)}\,,\\
\delta_2 &= \frac{ -R\, a\, b\ \sin\theta}{\sqrt{r^2+a^2}}\,\Bigl[\frac{r^2+a^2}{r^2+a^2 \cos^2\theta}\cos^2\theta \cos\phi\, \,d\phi\wedge d\psi + \sin\phi\, \frac{\cos\theta}{\sin\theta}\,
d\theta\wedge d\psi \Bigr]\,,\\
\omega &=  \frac{R\,a^2}{\sqrt{2}\,(r^2+a^2 \cos^2\theta)}\,(\sin^2\theta\,d\phi + \cos^2\theta\,d\psi)\,.
\end{align}
\end{subequations}
Expanding this solution up to order $b^2$, keeping $Q_1$, $Q_5$ and $R$ fixed, yields
 \begin{subequations}\label{eq:bkg2ndorder}
\allowdisplaybreaks
 \begin{align}
\frac{ds^2_6}{\sqrt{Q_1 Q_5}}&=V^{-2}\,\left[ds^2_{AdS_3}+ b^2 \,\delta g_{\mu\nu}\,dx^\mu dx^\nu\right]\nonumber\\
&+\left(1-\frac{b^2}{4\,a_0^2}\, B_+ B_-\,\sin^2\theta\right)\,d\theta^2
+\sin^2\theta\,\left(1+\frac{b^2}{4\,a_0^2}\, B_+ B_-\,\sin^2\theta\right)\,(d\hat\phi+b^2 A^\phi)^2\nonumber\\
&+\cos^2\theta\,\left(1-\frac{b^2}{4\,a_0^2}\, B_+  B_-\,(\cos^2\theta+1)\right)\,(d\hat\psi+b^2 A^\psi)^2\,,\\
C&=C_{0} +Q_5\,b^2 \left[\sin^2\theta\, d\hat \phi \wedge A^\psi +\cos^2\theta\, d\hat \psi \wedge A^\phi -\frac{B_+ B_-}{2\,a_0^2}\,\sin^2\theta\,\cos^2\theta\,d\hat \phi\wedge d\hat \psi\right]\,,\\
e^{2\phi_1}&=\frac{Q_1}{Q_5}\left[1+ \frac{b^2}{2\,a_0^2}\,\left[(B_+ Y^{++})^2 + (B_- Y^{--})^2 + B_+ B_- Y^{++} Y^{--}\right]\right]\,,\\
e^{2\phi_2}&=\frac{Q_5}{Q_1}\left[1+ \frac{b^2}{2\,a_0^2}\,B_+ B_- Y^{++} Y^{--} \right]\,,\\
\chi_1 &= \sqrt{\frac{Q_5}{Q_1}}\,\frac{b}{2\,a_0}\,(B_+ Y^{++} + B_- Y^{--})\,,\quad \chi_2 = \sqrt{\frac{Q_1}{Q_5}}\,\frac{b}{2\,a_0}\,(B_+ Y^{++} + B_- Y^{--})\,,\label{eq:chis}\\
\frac{B}{\sqrt{Q_1 Q_5}}& = \frac{b}{2\,a_0}\,(Y^{++} *_{AdS_3} d B_+ -  B_+ *_{S^3} d Y^{++}+Y^{--} *_{AdS_3} d B_- -  B_- *_{S^3} d Y^{--})\,,\label{eq:B}
\end{align}
\end{subequations}
where
\begin{equation}
V=1- \frac{b^2}{8\,a_0^2}\,B_+ B_-\,(\cos^2\theta+1)\,,
\end{equation}
and $B_\pm$, $Y^{\pm\pm}$, $\delta g_{\mu\nu}$, $A^\phi$ and $A^\psi$ are defined in (\ref{eq:Bpm}) and (\ref{eq:bkgcoefficients}).

\section{On $S^3$ spherical harmonics}
\label{sec:harmonics}

We embed the 3-sphere of unit radius in $\mathrm{R}^4$ as follows
\begin{align}
x^1+ix^2=\sin\theta e^{i\hat{\phi}},\quad\quad x^3+ix^4=\cos\theta e^{i\hat{\psi}}
\end{align}
with $\theta\in [0,\pi/2]$ and $\hat{\phi},\hat{\psi}\in [0,2\pi]$. The round metric of $S^3$ reads
\begin{align}\label{eq:s3rm}
ds^2_{S^3}=d \theta^2+\sin^2\theta\, d\hat{\phi}^2+\cos^2\theta\, d\hat{\psi}^2
\end{align}
and the volume form is
\begin{align}
\mathrm{vol}_{S^3}= \sin\theta\,\cos\theta\,d\theta\wedge d\hat \phi\wedge d\hat \psi~.
\end{align}   
\\
The scalar spherical harmonics of degree $k$ are defined as
\begin{align}
Y^{I_k}=x^{\left(i_1\right.}\cdots x^{\left.i_k\right)}~,\quad\quad i_k=1,\dots,4
\end{align} 
with symmetrized and traceless indices with respect the flat metric of $\mathrm{R}^4$. They satisfy the equation\footnote{We define the Laplace-de Rham operator operator acting on $p$-forms as $\Delta_{S^3} = -(d+\delta)^2$, where $\delta =(-1)^p * d *$ when it acts on a $p$-form; on scalars $\Delta_{S^3}$ reduces to the $\Box_{S^3}$ defined in the text. Our convention for the Hodge dual in $d$-dimensions is
\begin{equation}
* \omega^{(p)} = \frac{\sqrt{|g|}}{p! (d-p)!}{\epsilon_{i_1\ldots i_{d-p}}}^{j_1\ldots j_p}\omega^{(p)}_{j_1\ldots j_p}dx^{i_1}\wedge \ldots dx^{i_{d-p}}\,.
\end{equation}
}
\begin{align}
\Delta_{S^3}Y^{I_k}=-k(k+2)Y^{I_k}~
\end{align}
and enjoy the orthogonality relation
\begin{align}
\label{orto}
\int d\Omega_3 Y^{I_{k_1}}Y^{I_{k_2}}=\frac{2^{1-k}\pi^2}{\Gamma(k+2)}\delta^{I_{k_1},I_{k_1}}~.
\end{align}

The scalar spherical harmonics of degree 1, used in the text, are linear combinations of the ones defined above. In particular we have (see~\eqref{eq:Bpm})
\begin{align}\label{eq:BpmA}
Y^{++}=x^1 +i x^2,\quad\quad Y^{--}=x^1 -i x^2=\left(Y^{++}\right)^*
\end{align}
and using the relation \eqref{orto} we have 
\begin{align}
\int d\Omega_3 Y^{\pm\pm}Y^{\mp\mp}=\pi^2,\quad\quad \int d\Omega_3 Y^{\pm\pm}Y^{\pm\pm}=0~.
\end{align}   

We can also construct vector harmonics ${Y}^{(1)}_{I_k}$ simply by taking the product of a scalar harmonic and a 1-form: $Y^{\left[I_k\right.}dx^{\left.j\right]}$ where the index $j$ is antisymmetrized with the index $i_1$ in $I_k$. The vector harmonics satisfy
\begin{align}\label{eq:1kk2}
\Delta_{S^3}{Y}^{(1)}_{I_k}=-(1+k(k+2)){Y}^{(1)}_{I_k},\quad\quad 
d*_{S^3} {Y}^{(1)}_{I_k}=0~. 
\end{align}
However, in this work we need to keep track only of the contributions of the scalar harmonics and their derivatives $dY^{I_k}$.
It is easy to see that the $dY^{I_k}$'s satisfy the same equation as the scalar harmonics
\begin{align}
\Delta_{S^3} dY^{I_k}=-k(k+2) dY^{I_k}
\end{align}   
and so are clearly orthogonal to the vector harmonics which satisfy~\eqref{eq:1kk2}. For the particular combination~\eqref{eq:BpmA} we have
\begin{align}
\int *_{S^3}dY^{\pm\pm}\wedge dY^{\mp\mp}=3\pi^2,\quad\quad\int *_{S^3}dY^{\pm\pm}\wedge dY^{\pm\pm}=0~.
\end{align}
It is then straightforward to isolate the contributions proportional to $Y^{++}$ and $dY^{++}$ in~\eqref{eq:pert1} by using the orthogonality properties mentioned above
\begin{subequations}
  \label{eq:exY}
\begin{align}
  f(\theta,\phi,\psi) & =  Y^{++} \left[\frac{1}{\pi^2} \int d\Omega_3 \; Y^{--} f(\theta,\phi,\psi)\right] + \ldots~, \\
   v(\theta,\phi,\psi) & =  dY^{++} \left[\frac{1}{3\pi^2} \int d\Omega_3 \; *_{S^3} dY^{--} \wedge v(\theta,\phi,\psi)\right] + \ldots~,
\end{align}
\end{subequations}
where $f$ and $v$ are an arbitrary scalar function and a 1-form on $S^3$.

\section{Some properties of the $D$-integrals}
\label{sec:cftconv}

The unnormalized boundary-to-bulk propagator for a scalar field propagating in Euclidean AdS$_{d+1}$ is
\begin{equation}
  \label{eq:b2bd}
  K_{\Delta}(w,\vec{z}) = 
\left[\frac{w_0}{w_0^2 + (\vec{w}-\vec{z})^2} \right]^{\Delta}=\frac{
1}{\Gamma(\Delta)} \int_0^\infty \!\! dt\, w_0^{\Delta}\, t^{\Delta-1} e^{-t(w_0^2+(\vec{w}-\vec{z})^2)}\;,
\end{equation}
where $\Delta$ is the conformal dimension of the dual operator (related to the mass as usual: $m^2=\Delta (\Delta -d)$). The $D$-integrals arising from a $4$-point contact vertex in the bulk take the following form
\begin{equation}
  \label{eq:Dintg}
  D_{\Delta_1 \Delta_2 \Delta_3 \Delta_4} = \int d^{d+1}w \, \sqrt{\bar g}\, \prod_{i=1}^4 K_{\Delta_i}(w,\vec{z}_i)\;,
\end{equation}
where the AdS$_{d+1}$ metric in the Euclidean Poincar\'e coordinates is
\begin{equation}
  \label{eq:AdSdPc}
  d{\bar s}^2 = \frac{1}{w_0^2} \left(dw_0^2 + \sum_{i=1}^d dw_i^2 \right)~.
\end{equation}
By using the representation of the propagator in terms of Schwinger parameters given in~\eqref{eq:b2bd} it is straightforward to perform the integration over the interaction point $(w_0,\vec{w})$ and obtain
\begin{equation}
  \label{eq:Dintg2}
   D_{\Delta_1 \Delta_2 \Delta_3 \Delta_4} = \Gamma\left(\frac{\hat{\Delta}-d}{2}\right)\int_0^\infty \prod_i\left[ dt_i \frac{t_i^{\Delta_i-1}}{\Gamma(\Delta_i)} \right] \frac{\pi^{d/2}}{2 T^{\frac{\hat\Delta}{2}}} e^{-\sum_{i,j=1}^4 |z_{ij}|^2 \frac{t_i t_j}{2 T}}\,
\end{equation}
with $T= \sum_i t_i$, $\hat\Delta=\sum_i \Delta_i$ and ${z}_{ij}={z}_i - {z}_j$, where $z_j$ are the standard complex coordinates for a plane $z_j=x_j+i y_j$. Since in this paper we are interested in the AdS$_3$ case, we set $d=2$. Once written in terms of Schwinger parameter, one can see that $D_{1111}$ is proportional to the massless box-integral in four dimensions with external massive state the result can be written in term of logarithms and dilogarithms
\begin{equation}
  \label{eq:D1111}
  D_{1111}= 
\frac{ \pi}{2 |z_{13}|^2 |z_{24}|^2 (z-\bar{z})} \left[2{\rm Li}_2(z) - 2 {\rm Li}_2(\bar{z}) + \ln(z \bar{z}) \ln\frac{1-z}{1-\bar{z}}\right]~,
\end{equation}
where $z$ is the crossratio defined in~\eqref{eq:pirat}.

The result in~\eqref{eq:Dintg2} is proportional to the Bloch-Wigner dilogarithm $D(z,\bar{z})$
\begin{equation}
  \label{eq:D1111bis}
   D_{1111}= 
\frac{2\pi i}{|z_{13}|^2 |z_{24}|^2 (z-\bar{z})} D(z,\bar{z})\,
\end{equation}
where
\begin{equation}
  \label{eq:BWD}
  \begin{aligned}
  D(z,\bar{z}) & =  {\rm Im} [{\rm Li}_2(z)] + {\rm Arg}[\ln(1-z)] \ln|z| 
 \\ & =  \frac{1}{2i} \left[{{\rm Li}_2(z)-{\rm Li}_2(\bar{z})+ \frac{1}{2} \ln(z\bar{z}) \ln\frac{1-z}{1-\bar{z}} }\right]\,.
  \end{aligned}
\end{equation}
The function $D(z,\bar{z})$ is a real-analytic function\footnote{In order for this to hold,  $\bar{z}$ has to be the complex conjugate of $z$, so the correlator~\eqref{eq:D1111bis} has a more complicated analytic structure in Minkowski space where $\bar{z}\not=z^*$, see~\cite{Maldacena:2015iua} and references therein: in that case it has non-trivial monodromies around $z=1$ and $z=\infty$ with $\bar{z}$ fixed (and vice versa).} except in $z=0,1$. It is continuous also in those two points, but not differentiable (since it has singularities of the type $y\log(y)$, where $y={\rm Im}[z]$ or $y={\rm Im}[1-z]$ and $y\to 0$). Moreover we have the following useful identities
\begin{equation}
  \label{eq:BWiden}
  D(z,\bar{z}) = - D\left(\frac{1}{z},\frac{1}{\bar{z}}\right) =  
  - D\left({1-z},{1-\bar{z}}\right) ~,
\end{equation}
which implies
\begin{equation}
  \label{eq:BWiden2}
 D(z,\bar{z}) = D\left(1-\frac{1}{z},1-\frac{1}{\bar{z}}\right) =  
  D\left(\frac{1}{1-z},\frac{1}{1-\bar{z}}\right) =  
  - D\left(\frac{-z}{1-z},\frac{-\bar{z}}{1-\bar{z}}\right) ~.
\end{equation}

Our correlator involves also the D-integrals of the type $D_{1122}$ and permutations, {\rm i.e.} we have two $\Delta_i$ equal to two and the other two equal to one. By using the expression in terms of Schwinger parameters~\eqref{eq:Dintg2} it is easy to write these integrals as derivatives of $D_{1111}$. The D-integrals of this type that are relevant for our correlator are
\begin{equation}
  \label{eq:D2211der}
   D_{2211} = -\frac{\partial D_{1111}}{\partial |z_{12}|^2} \;,
 \end{equation}
and its permutations.

The D-functions that appear in the gravity computation are evaluated at the points $0,\infty,1,z$ and are denoted by $\hat{D}_{\Delta_1,\Delta_2,\Delta_3,\Delta_4}$ (see eq. (\ref{eq:Dhat})).  They can be obtained as a limit and are given by
\begin{equation}\label{eq:D2211}
\begin{aligned}
\hat D_{2211}(z,\bar z)&= \lim_{z_2\to\infty} |z_2|^{4} D_{2211}(z_1=0,z_2,z_3=1,z_4=z)
\\ = &
-\frac{2\pi i\,|1-z|^2}{(z-\bar z)^2}\left[\frac{z+\bar z}{z-\bar z} D(z,\bar z)+\frac{\log|1-z|^2}{2i}+\frac{z+\bar z-2 z\bar z}{4i|1-z|^2}\log|z|^2\right]\,.
\end{aligned}
\end{equation}
Some useful relations between the $\hat{D}$'s appearing in the intermediate results of Appendix~\ref{sec:red2dintegrals} are
\begin{subequations}
\begin{align}
\hat D_{1122}(z,\bar z)&=\frac{\hat D_{2211}(z,\bar z)}{|1-z|^2}\,,\quad \hat D_{1212}(z,\bar z)=\hat D_{2121}(z,\bar z)= \frac{1}{|z|^2} \hat D_{2211}\left(\frac{z-1}{z},\frac{\bar z-1}{\bar z}\right)\,,\\
\hat D_{1221}(z,\bar z)&=|z|^2\,\hat D_{2112}(z,\bar z)=\hat D_{2211}(1-z,1-\bar z)\,.
\end{align}
\end{subequations}
The symmetry under exchange of the first two points in $D_{2211}$ implies
\begin{equation}
\label{eq:hatDsymmetry}
\hat D_{2211}\left(\frac{1}{z},\frac{1}{\bar z}\right) = |z|^2 \,\hat D_{2211}(z,\bar z)\,.
\end{equation}
All the $\hat D$ functions are linear combinations of $D(z,\bar z)$, $\log|1-z|^2$ and $\log|z|^2$ with coefficients that are ratios of polynomials in $z$ and $\bar z$. Inverting the relations between the functions $\hat D$ and $D(z,\bar z)$, $\log|1-z|^2$, $\log|z|^2$ one finds the useful identities
\begin{equation}\label{eq:Dhatid}
\hat D_{1111}=\hat D_{2211}+\hat D_{1221}+\hat D_{2121}\,,\quad \pi \log|z|=|1-z|^2 (\hat D_{1221}-\hat D_{2121})+(|z|^2-1)\,\hat D_{2211}\,.
\end{equation}

\section{Reduction to $D$-integrals}
\label{sec:red2dintegrals}

 In this appendix we derive our main result (\ref{eq:finalcorrelator}) starting from (\ref{eq:correlatorsemifinalcylinder}) and (\ref{eq:Js}). We first work in Euclidean global coordinates and, for notational coherence, we set 
 \begin{equation}
B_0(\mathbf{r}'_e|t_e, y)\equiv K_1^\mathrm{Glob}(\mathbf{r}'_e|t_e, y)=\frac{1}{2}\,\frac{a_0}{\sqrt{r'^2+a_0^2} \,\cosh ((t'_e-t_e)/R) - r'\,\cos ((y'-y)/R)}\,;
\end{equation}
with this notation eq. (\ref{eq:B0}) reads $B_0(\mathbf{r}'_e)=B_0(\mathbf{r}'_e|0, 0)$. We can rewrite the r.h.s. of eq. (\ref{eq:correlatorsemifinalcylinder}), with the source given in (\ref{eq:Js}), as
\begin{equation}\label{eq:sumI}
- \frac{b^2 e^{-\frac{t_e}{R}}}{8\pi} \left[-\frac{1}{3}I_1+\frac{1}{3}I_2-\frac{5}{3}I_3+\frac{7}{3}I_4-4 I_5 +4 I_6\right]\,.
\end{equation}
The integrals $I_i$ are defined as
\begin{subequations}
\begin{align}
I_1 &= \int d^3 \mathbf{r}'_e\,\sqrt{\bar g}\,B_{0}(\mathbf{r}'_e|t_e,y)\,B_{0}(\mathbf{r}'_e|0,0)\,B_{+}(\mathbf{r}'_e)\,B_{-}(\mathbf{r}'_e)\,,\\
I_2 &=  \int d^3 \mathbf{r}'_e\,\sqrt{\bar g}\,B_{0}(\mathbf{r}'_e|t_e,y)\,B_{0}(\mathbf{r}'_e|0,0)\,\partial'_\mu B_{+}(\mathbf{r}'_e)\,\partial'^\mu B_{-}(\mathbf{r}'_e)\,,\\
I_3 &=  \int d^3 \mathbf{r}'_e\,\sqrt{\bar g}\,B_{0}(\mathbf{r}'_e|t_e,y)\,\partial'^\mu B_{0}(\mathbf{r}'_e|0,0)\,B_{-}(\mathbf{r}'_e)\,\partial'_\mu B_{+}(\mathbf{r}'_e)\,,\\
I_4 &=  \int d^3 \mathbf{r}'_e\,\sqrt{\bar g}\,B_{0}(\mathbf{r}'_e|t_e,y)\,\partial'^\mu B_{0}(\mathbf{r}'_e|0,0)\,B_{+}(\mathbf{r}'_e)\,\partial'_\mu B_{-}(\mathbf{r}'_e)\,,\\
I_5 &=  \int d^3 \mathbf{r}'_e\,\sqrt{\bar g}\,B_{0}(\mathbf{r}'_e|t_e,y)\,R^2 \partial^2_{t'_e} B_{0}(\mathbf{r}'_e|0,0)\,\frac{a_0^4}{(r'^2+a_0^2)^2}\,,\\
I_6 &=  \int d^3 \mathbf{r}'_e\,\sqrt{\bar g}\,B_{0}(\mathbf{r}'_e|t_e,y)\,R\, \partial_{t'_e} B_{0}(\mathbf{r}'_e|0,0)\,\frac{a_0^2}{r'^2+a_0^2}\,,\\
I_7 &=  \int d^3 \mathbf{r}'_e\,\sqrt{\bar g}\,B_{0}(\mathbf{r}'_e|t_e,y)\,i\,R\, \partial_{y'} B_{0}(\mathbf{r}'_e|0,0)\,\frac{a_0^2}{r'^2+a_0^2}\,,
\end{align}
\end{subequations}
where in the last line we have also introduced the integral $I_7$ for later convenience.
Our goal is to rewrite the integrals $I_i$ in terms of the D-functions evaluated at the boundary points $0,\infty,1,z$ as defined in (\ref{eq:Dhat}). We can use the identities induced by the change of coordinates (\ref{eq:glob2poinc}):
\begin{equation}
\begin{aligned}
&B_{0}(\mathbf{r}'_e|t_e,y) = |z| K_1(\mathbf{w}|z,\bar z)\,,\\
&B_+(\mathbf{r}'_e) = \lim_{z\to \infty}|z|^2 K_1(\mathbf{w}|z,\bar z)\equiv K_1(\mathbf{w}|\infty)=w_0\,,\quad B_-(\mathbf{r}'_e)=K_1(\mathbf{w}|0)\,.
\end{aligned}
\end{equation}
Then one immediately finds
\begin{equation}
|z|^{-1}I_1 = \hat D_{1111}\,.
\end{equation}
The relation
\begin{equation}
\partial'_\mu B_{+}(\mathbf{r}'_e)\,\partial'^\mu B_{-}(\mathbf{r}'_e) = \frac{a_0^2}{r'^2+a_0^2}-\frac{2\,a_0^4}{(r'^2+a_0^2)^2} = K_1(\mathbf{w}|\infty)K_1(\mathbf{w}|0)-2\, K_2(\mathbf{w}|\infty)K_2(\mathbf{w}|0)
\end{equation}
yields
\begin{equation}
|z|^{-1}I_2= \hat D_{1111}-2 \hat D_{2211}\,.
\end{equation}
$I_3$ can be computed by explicitly writing the integral in Poincar\'e coordinates 
\begin{equation}\label{eq:I3}
\begin{aligned}
|z|^{-1}I_3&=\int d^3 \mathbf{w}\,w_0^{-1}\,\frac{w_0}{w_0^2+|w-z|^2}\,\partial_{w_0}\left(\frac{w_0}{w_0^2+|w-1|^2}\right)\,\frac{w_0}{w_0^2+|z|^2}\\
&=\int d^3 \mathbf{w}\,\frac{w_0}{w_0^2+|w-z|^2}\,\left(\frac{1}{w_0^2+|w-1|^2}-\frac{2\,w_0^2}{(w_0^2+|w-1|^2)^2}\right)\,\frac{1}{w_0^2+|z|^2}\\
&=\hat D_{1111}-2\,\hat D_{1221}\,.
\end{aligned}
\end{equation}
Moreover the fact that $B_{0}(\mathbf{r}'_e|t_e,y)$ is an even function of $t'_e-t_e$ and $y'-y$ implies that
\begin{equation}
\begin{aligned}
&\int d^3 \mathbf{r}'_e\,\sqrt{\bar g}\,B_{0}(\mathbf{r}'_e|t_e,y)\,\partial'^\mu B_{0}(\mathbf{r}'_e|0,0)\,B_{-}(\mathbf{r}'_e)\,\partial'_\mu B_{+}(\mathbf{r}'_e)=\\
&\qquad =  \int d^3 \mathbf{r}'_e\,\sqrt{\bar g}\,B_{0}(\mathbf{r}'_e|0,0)\,\partial'^\mu B_{0}(\mathbf{r}'_e|t_e,y)\,B_{+}(\mathbf{r}'_e)\,\partial'_\mu B_{-}(\mathbf{r}'_e)\,,
\end{aligned}
\end{equation}
since the change of integration variables $(t'_e,y')\to (-t'_e, -y')$ followed by $(t'_e,y')\to (t'_e-t_e, y'-y)$ exchanges $B_{0}(\mathbf{r}'_e|t_e,y)$ with $B_{0}(\mathbf{r}'_e|0,0)$ and $B_{+}(\mathbf{r}'_e)$ with $B_{-}(\mathbf{r}'_e)$. Then
\begin{equation}
\begin{aligned}\label{eq:I3I4}
\frac{I_3+I_4}{|z|}&=\frac{1}{2}\int \!d^3 \mathbf{w}\,\sqrt{\bar g}\,\partial^\mu [K_1(w|z) K_1(w|1)][K_1(w|0)\partial_\mu K_1(w|\infty)+K_1(w|\infty)\partial_\mu K_1(w|0)]\\
&=-\frac{I_2}{|z|} + \hat D_{1111} = 2\,\hat D_{2211}\,,
\end{aligned}
\end{equation}
where in the last line we have integrated by parts and used $\Box K_1+ K_1=0$. From (\ref{eq:I3}) and (\ref{eq:I3I4}) we then deduce $I_4$
\begin{equation}
|z|^{-1}I_4=-\hat D_{1111}+2\,\hat D_{2211} +2\,\hat D_{1221}\,.
\end{equation}
To compute $I_5$ one notes that
\begin{equation}
\frac{a_0^4\,R}{(r'^2+a_0^2)^2}\, \partial_{t'_e} B_{0}(\mathbf{r}'_e|0,0)=\frac{1}{2}[B_{-}(\mathbf{r}'_e)\,\partial'_\mu B_{+}(\mathbf{r}'_e)-B_{+}(\mathbf{r}'_e)\,\partial'_\mu B_{-}(\mathbf{r}'_e)]\partial'^\mu B_{0}(\mathbf{r}'_e|0,0)\,,
\end{equation}
which implies
\begin{equation}\label{eq:relationforI5}
\begin{aligned}
&\int d^3 \mathbf{r}'_e\,\sqrt{\bar g}\,B_{0}(\mathbf{r}'_e|t_e,y)\,R\, \partial_{t'_e} B_{0}(\mathbf{r}'_e|0,0)\,\frac{a_0^4}{(r'^2+a_0^2)^2}=\frac{1}{2}(I_3-I_4) =\\
&= |z| (\hat D_{1111}-2\,\hat D_{1221}-\hat D_{2211})\,.
\end{aligned}
\end{equation}
Then
\begin{equation}
\begin{aligned}
I_5 &=-\int d^3 \mathbf{r}'_e\,\sqrt{\bar g}\,R\,\partial_{t'_e}B_{0}(\mathbf{r}'_e|t_e,y)\,R\, \partial_{t'_e} B_{0}(\mathbf{r}'_e|0,0)\,\frac{a_0^4}{(r'^2+a_0^2)^2}\\
&=R\, \partial_{t_e}\int d^3 \mathbf{r}'_e\,\sqrt{\bar g}\,B_{0}(\mathbf{r}'_e|t_e,y)\,R\, \partial_{t'_e} B_{0}(\mathbf{r}'_e|0,0)\,\frac{a_0^4}{(r'^2+a_0^2)^2}\\
&=(z \partial_z+{\bar z}\partial_{\bar z})\, (|z| (\hat D_{1111}-2\,\hat D_{1221}-\hat D_{2211}))\\
&=|z|\,\left(2\,\hat D_{1122}+\hat D_{2121}-\hat D_{1221}-\frac{\pi}{|1-z|^2} (1-\log|z|)\right)\,,
\end{aligned}
\end{equation}
where we have first integrated by parts, exploited the fact that $B_{0}(\mathbf{r}'_e|t_e,y)$ is a function of $t'_e-t_e$ and then used (\ref{eq:relationforI5}); the last line follows by substituting the explicit expressions for the functions $\hat D$ given in Appendix \ref{sec:cftconv}. Finally $I_6$ and $I_7$ follow from similar manipulations
\begin{equation}
\begin{aligned}
I_6&=- \int d^3 \mathbf{r}'_e\,\sqrt{\bar g}\,R\, \partial_{t'_e} B_{0}(\mathbf{r}'_e|t_e,y)\,B_{0}(\mathbf{r}'_e|0,0)\,\frac{a_0^2}{r'^2+a_0^2}\\
&= R\, \partial_{t_e}\int d^3 \mathbf{r}'_e\,\sqrt{\bar g}\,B_{0}(\mathbf{r}'_e|t_e,y)\,B_{0}(\mathbf{r}'_e|0,0)\,\frac{a_0^2}{r'^2+a_0^2}\\
&=(z \partial_z+{\bar z}\partial_{\bar z})\, (|z| \,\hat D_{1111})=-\frac{\pi\,|z|}{|1-z|^2} \,\log|z|\,,
\end{aligned}
\end{equation}
\begin{equation}
\begin{aligned}
I_7&=- \int d^3 \mathbf{r}'_e\,\sqrt{\bar g}\,i\,R\, \partial_{y'} B_{0}(\mathbf{r}'_e|t_e,y)\,B_{0}(\mathbf{r}'_e|0,0)\,\frac{a_0^2}{r'^2+a_0^2}\\
&= i\,R\, \partial_{y}\int d^3 \mathbf{r}'_e\,\sqrt{\bar g}\,B_{0}(\mathbf{r}'_e|t_e,y)\,B_{0}(\mathbf{r}'_e|0,0)\,\frac{a_0^2}{r'^2+a_0^2}\\
&=-(z \partial_z-{\bar z}\partial_{\bar z})\, (|z| \,\hat D_{1111})=-|z|\,\frac{z-\bar z}{|1-z|^2}\,\hat D_{2211}\,,
\end{aligned}
\end{equation}
where the last steps we have used (\ref{eq:D1111bis}) and (\ref{eq:D2211}).

Substituting the simplified expressions for the integrals $I_i$ in (\ref{eq:sumI}), using the identities (\ref{eq:Dhatid}) and performing the transformation from the cylinder to the plane correlator yields the result (\ref{eq:finalcorrelator}).

\section{The correlator with $O_L= O^{+-}$}
\label{sec:Opmap}

It is straightforward to repeat our computations to derive a different 4-point correlator where the light operator $O^{++}$ is replaced by $O^{+-}$, while the heavy operators are left unchanged. In this appendix, by a slight abuse of notation, we will denote $O_L = O^{+-}$ and $\mathcal{G}$ will represent the correlator (\ref{eq:corrHHLL}) for this choice of $O_L$.

The methods of Appendix \ref{sec:free-corr} give the orbifold point result
\begin{equation}
\mathcal{G}(z,\bar z)=\sqrt{\frac{\bar z}{z}} + \frac{|B|^2}{2 N}\sqrt{\frac{\bar z}{z}} \,\left[ z+\frac{1}{\bar z}-2+\frac{|1-z|^2}{\bar z}\right]\,.
\end{equation}
Note that the term of order $N$ is now absent.

The gravity computation follows the lines of Section \ref{sec:4pointcomp}, where now one looks for a perturbation of the form (\ref{eq:pert++}) with the spherical harmonic $Y^{++}$ replaced by $Y^{+-}$. The r.h.s. of eq. (\ref{eq:correlatorsemifinalcylinder}) becomes
\begin{equation}\label{eq:sumIbis}
- \frac{b^2 e^{-i\frac{y}{R}}}{8\pi} \left[\frac{1}{3}I_1-\frac{1}{3}I_2+\frac{5}{3}I_3+\frac{5}{3}I_4-4 I_5 +4 I_7\right]= b^2 \frac{|z|}{|1-z|^2}\,\sqrt{\frac{\bar z}{z}}\,\left[\frac{z}{\pi} \,\hat D_{2211}-\frac{1}{2}\right]\,,
\end{equation}
having used the results of Appendix \ref{sec:red2dintegrals} and the second identity in (\ref{eq:Dhatid}). The gravity correlator with $O_L=O^{+-}$ up to order $b^2$ is then
\begin{equation} \label{eq:finalcorrelatorab+-}
\mathcal{G}_\mathrm{grav}(z,\bar z) = \sqrt{\frac{\bar z}{z}}\,\left[1+\frac{b^2}{a_0^2}\,\left(\frac{z}{\pi}\,\hat D_{2211}-\frac{1}{2}\right)\right]\,.
\end{equation}

\providecommand{\href}[2]{#2}\begingroup\raggedright\endgroup


\begin{thebibliography}{10}

\bibitem{Maldacena:1997re}
J.~M. Maldacena, ``{The large N limit of superconformal field theories and
  supergravity},'' {\em Adv. Theor. Math. Phys.} {\bf 2} (1998) 231--252,
\href{http://arXiv.org/abs/hep-th/9711200}{{\tt hep-th/9711200}}.

\bibitem{Gubser:1998bc}
S.~S. Gubser, I.~R. Klebanov, and A.~M. Polyakov, ``{Gauge theory correlators
  from non-critical string theory},'' {\em Phys. Lett.} {\bf B428} (1998)
  105--114,
\href{http://arXiv.org/abs/hep-th/9802109}{{\tt hep-th/9802109}}.

\bibitem{Witten:1998qj}
E.~Witten, ``{Anti-de Sitter space and holography},'' {\em Adv. Theor. Math.
  Phys.} {\bf 2} (1998) 253--291,
\href{http://arXiv.org/abs/hep-th/9802150}{{\tt hep-th/9802150}}.

\bibitem{Baggio:2012rr}
M.~Baggio, J.~de~Boer, and K.~Papadodimas, ``{A non-renormalization theorem for
  chiral primary 3-point functions},''
  \href{http://dx.doi.org/10.1007/JHEP07(2012)137}{{\em JHEP} {\bfseries 1207}
  (2012) 137},
\href{http://arxiv.org/abs/1203.1036}{{\ttfamily arXiv:1203.1036 [hep-th]}}.

\bibitem{Kanitscheider:2006zf}
I.~Kanitscheider, K.~Skenderis, and M.~Taylor, ``{Holographic anatomy of
  fuzzballs},'' {\em JHEP} {\bf 0704} (2007) 023,
\href{http://arXiv.org/abs/hep-th/0611171}{{\tt hep-th/0611171}}.

\bibitem{Kanitscheider:2007wq}
I.~Kanitscheider, K.~Skenderis, and M.~Taylor, ``{Fuzzballs with internal
  excitations},'' {\em JHEP} {\bf 06} (2007) 056,
\href{http://arXiv.org/abs/0704.0690}{{\tt 0704.0690}}.

\bibitem{Giusto:2015dfa}
S.~Giusto, E.~Moscato, and R.~Russo, ``{AdS$_{3}$ holography for 1/4 and 1/8
  BPS geometries},'' {\em JHEP} {\bf 11} (2015) 004,
\href{http://arXiv.org/abs/1507.00945}{{\tt 1507.00945}}.

\bibitem{Rastelli:2016nze}
L.~Rastelli and X.~Zhou, ``{Mellin amplitudes for $AdS_5\times S^5$},'' {\em
  Phys. Rev. Lett.} {\bf 118} (2017), no.~9, 091602,
\href{http://arXiv.org/abs/1608.06624}{{\tt 1608.06624}}.

\bibitem{DHoker:1999mqo}
E.~D'Hoker, D.~Z. Freedman, and L.~Rastelli, ``{AdS / CFT four point functions:
  How to succeed at z integrals without really trying},'' {\em Nucl. Phys.}
  {\bf B562} (1999) 395--411,
\href{http://arXiv.org/abs/hep-th/9905049}{{\tt hep-th/9905049}}.

\bibitem{Balasubramanian:2005qu}
V.~Balasubramanian, P.~Kraus, and M.~Shigemori, ``{Massless black holes and
  black rings as effective geometries of the D1-D5 system},'' {\em Class.
  Quant. Grav.} {\bf 22} (2005) 4803--4838,
\href{http://arXiv.org/abs/hep-th/0508110}{{\tt hep-th/0508110}}.

\bibitem{Lunin:2012gz}
O.~Lunin and S.~D. Mathur, ``{A toy black hole S-matrix in the D1-D5 CFT},''
  {\em JHEP} {\bf 02} (2013) 083,
\href{http://arXiv.org/abs/1211.5830}{{\tt 1211.5830}}.

\bibitem{Galliani:2016cai}
A.~Galliani, S.~Giusto, E.~Moscato, and R.~Russo, ``{Correlators at large c
  without information loss},'' {\em JHEP} {\bf 09} (2016) 065,
\href{http://arXiv.org/abs/1606.01119}{{\tt 1606.01119}}.

\bibitem{Balasubramanian:2016ids}
V.~Balasubramanian, B.~Craps, B.~Czech, and G.~Sárosi, ``{Echoes of chaos from
  string theory black holes},'' {\em JHEP} {\bf 03} (2017) 154,
\href{http://arXiv.org/abs/1612.04334}{{\tt 1612.04334}}.

\bibitem{Fitzpatrick:2015zha}
A.~L. Fitzpatrick, J.~Kaplan, and M.~T. Walters, ``{Virasoro Conformal Blocks
  and Thermality from Classical Background Fields},''
\href{http://arXiv.org/abs/1501.05315}{{\tt 1501.05315}}.

\bibitem{Hijano:2015zsa}
E.~Hijano, P.~Kraus, E.~Perlmutter, and R.~Snively, ``{Witten Diagrams
  Revisited: The AdS Geometry of Conformal Blocks},'' {\em JHEP} {\bf 01}
  (2016) 146,
\href{http://arXiv.org/abs/1508.00501}{{\tt 1508.00501}}.

\bibitem{Hijano:2015qja}
E.~Hijano, P.~Kraus, E.~Perlmutter, and R.~Snively, ``{Semiclassical Virasoro
  blocks from AdS$_{3}$ gravity},'' {\em JHEP} {\bf 12} (2015) 077,
\href{http://arXiv.org/abs/1508.04987}{{\tt 1508.04987}}.

\bibitem{daCunha:2016crm}
B.~C. da~Cunha and M.~Guica, ``{Exploring the BTZ bulk with boundary conformal
  blocks},''
\href{http://arXiv.org/abs/1604.07383}{{\tt 1604.07383}}.

\bibitem{Zarembo:2010rr} 
  K.~Zarembo,
  JHEP {\bf 1009}, 030 (2010)
  doi:10.1007/JHEP09(2010)030
  [arXiv:1008.1059 [hep-th]].
  
  \bibitem{Costa:2010rz} 
  M.~S.~Costa, R.~Monteiro, J.~E.~Santos and D.~Zoakos,
  JHEP {\bf 1011}, 141 (2010)
  doi:10.1007/JHEP11(2010)141
  [arXiv:1008.1070 [hep-th]].

\bibitem{Buchbinder:2010ek}
  E.~I.~Buchbinder and A.~A.~Tseytlin,
  ``Semiclassical four-point functions in AdS$_5 \times S^5$,''
  JHEP {\bf 1102} (2011) 072,
\href{http://arXiv.org/abs/1012.3740}{{\tt 1012.3740}}.
  
\bibitem{Sfondrini:2014via}
  A.~Sfondrini,
  J.\ Phys.\ A {\bf 48} (2015) no.2,  023001
  doi:10.1088/1751-8113/48/2/023001
  [arXiv:1406.2971 [hep-th]].

\bibitem{Sen:1995in}
A.~Sen, ``{Extremal black holes and elementary string states},'' {\em Mod.
  Phys. Lett.} {\bf A10} (1995) 2081--2094,
\href{http://arXiv.org/abs/hep-th/9504147}{{\tt hep-th/9504147}}.

\bibitem{Strominger:1996sh}
A.~Strominger and C.~Vafa, ``{Microscopic origin of the Bekenstein-Hawking
  entropy},'' {\em Phys.Lett.} {\bf B379} (1996) 99--104,
  \href{http://arXiv.org/abs/hep-th/9601029}{{\tt hep-th/9601029}}.

\bibitem{Lunin:2001jy}
O.~Lunin and S.~D. Mathur, ``{AdS/CFT duality and the black hole information
  paradox},'' {\em Nucl. Phys.} {\bf B623} (2002) 342--394,
\href{http://arXiv.org/abs/hep-th/0109154}{{\tt hep-th/0109154}}.

\bibitem{Lunin:2002iz}
O.~Lunin, J.~M. Maldacena, and L.~Maoz, ``{Gravity solutions for the D1-D5
  system with angular momentum},''
\href{http://arXiv.org/abs/hep-th/0212210}{{\tt hep-th/0212210}}.

\bibitem{Deger:1998nm}
S.~Deger, A.~Kaya, E.~Sezgin, and P.~Sundell, ``{Spectrum of D = 6, N=4b
  supergravity on AdS in three-dimensions x S**3},'' {\em Nucl. Phys.} {\bf
  B536} (1998) 110--140,
\href{http://arXiv.org/abs/hep-th/9804166}{{\tt hep-th/9804166}}.

\bibitem{D'Hoker:1999jp}
E.~D'Hoker, S.~D. Mathur, A.~Matusis, and L.~Rastelli, ``{The operator product
  expansion of N = 4 SYM and the 4- point functions of supergravity},'' {\em
  Nucl. Phys.} {\bf B589} (2000) 38--74,
\href{http://arXiv.org/abs/hep-th/9911222}{{\tt hep-th/9911222}}.

\bibitem{Arutyunov:2000ku}
G.~Arutyunov, S.~Frolov, and A.~C. Petkou, ``{Operator product expansion of the
  lowest weight CPOs in $\mathcal N=4$ SYM$_4$ at strong coupling},'' {\em
  Nucl. Phys.} {\bf B586} (2000) 547--588,
  \href{http://arXiv.org/abs/hep-th/0005182}{{\tt hep-th/0005182}}.
[Erratum: Nucl. Phys.B609,539(2001)].


\bibitem{Balasubramanian:2017fan}
V.~Balasubramanian, A.~Bernamonti, B.~Craps, T.~De~Jonckheere, and F.~Galli,
  ``{Heavy-Heavy-Light-Light correlators in Liouville theory},''
\href{http://arXiv.org/abs/1705.08004}{{\tt 1705.08004}}.

\bibitem{Avery:2010qw}
S.~G. Avery, ``{Using the D1D5 CFT to Understand Black Holes},''
\href{http://arXiv.org/abs/1012.0072}{{\tt 1012.0072}}.

\bibitem{Skenderis:2006ah}
K.~Skenderis and M.~Taylor, ``{Fuzzball solutions and D1-D5 microstates},''
  {\em Phys.Rev.Lett.} {\bf 98} (2007) 071601,
\href{http://arXiv.org/abs/hep-th/0609154}{{\tt hep-th/0609154}}.

\bibitem{Roy:2016zzv}
P.~Roy, Y.~K. Srivastava, and A.~Virmani, ``{Hair on non-extremal D1-D5 bound
  states},'' {\em JHEP} {\bf 09} (2016) 145,
\href{http://arXiv.org/abs/1607.05405}{{\tt 1607.05405}}.

\bibitem{Giusto:2013rxa}
S.~Giusto, L.~Martucci, M.~Petrini, and R.~Russo, ``{6D microstate geometries
  from 10D structures},'' {\em Nucl.Phys.} {\bf B876} (2013) 509--555,
\href{http://arXiv.org/abs/1306.1745}{{\tt 1306.1745}}.

\bibitem{DHoker:2002nbb}
E.~D'Hoker and D.~Z. Freedman, ``{Supersymmetric gauge theories and the AdS /
  CFT correspondence},''
\href{http://arXiv.org/abs/hep-th/0201253}{{\tt hep-th/0201253}}.

\bibitem{Liu:1998ty}
  H.~Liu and A.~A.~Tseytlin,
  ``On four point functions in the CFT / AdS correspondence,''
  Phys.\ Rev.\ D {\bf 59} (1999) 086002,
\href{http://arXiv.org/abs/hep-th/9807097}{{\tt hep-th/9807097}}.

\bibitem{Freedman:1998bj}
D.~Z. Freedman, S.~D. Mathur, A.~Matusis, and L.~Rastelli, ``{Comments on
  4-point functions in the CFT/AdS correspondence},'' {\em Phys. Lett.} {\bf
  B452} (1999) 61--68,
\href{http://arXiv.org/abs/hep-th/9808006}{{\tt hep-th/9808006}}.

\bibitem{D'Hoker:1999pj}
E.~D'Hoker, D.~Z. Freedman, S.~D. Mathur, A.~Matusis, and L.~Rastelli,
  ``{Graviton exchange and complete 4-point functions in the AdS/CFT
  correspondence},'' {\em Nucl. Phys.} {\bf B562} (1999) 353--394,
\href{http://arXiv.org/abs/hep-th/9903196}{{\tt hep-th/9903196}}.

\bibitem{Pakman:2009mi}
  A.~Pakman, L.~Rastelli and S.~S.~Razamat,
  JHEP {\bf 1005}, 099 (2010)
  doi:10.1007/JHEP05(2010)099
  [arXiv:0912.0959 [hep-th]].

\bibitem{Carson:2016uwf}
  Z.~Carson, S.~Hampton and S.~D.~Mathur,
  arXiv:1612.03886 [hep-th].
  
\bibitem{Fitzpatrick:2011hh}
A.~L. Fitzpatrick and D.~Shih, ``{Anomalous Dimensions of Non-Chiral Operators
  from AdS/CFT},'' {\em JHEP} {\bf 10} (2011) 113,
\href{http://arXiv.org/abs/1104.5013}{{\tt 1104.5013}}.

\bibitem{Alday:2017gde}
L.~F. Alday, A.~Bissi, and E.~Perlmutter, ``{Holographic Reconstruction of AdS
  Exchanges from Crossing Symmetry},''
\href{http://arXiv.org/abs/1705.02318}{{\tt 1705.02318}}.

\bibitem{Fitzpatrick:2016ive}
A.~L. Fitzpatrick, J.~Kaplan, D.~Li, and J.~Wang, ``{On information loss in
  AdS$_{3}$/CFT$_{2}$},'' {\em JHEP} {\bf 05} (2016) 109,
\href{http://arXiv.org/abs/1603.08925}{{\tt 1603.08925}}.

\bibitem{Fitzpatrick:2016mjq}
A.~L. Fitzpatrick and J.~Kaplan, ``{On the Late-Time Behavior of Virasoro
  Blocks and a Classification of Semiclassical Saddles},'' {\em JHEP} {\bf 04}
  (2017) 072,
\href{http://arXiv.org/abs/1609.07153}{{\tt 1609.07153}}.

\bibitem{Giusto:2013bda}
S.~Giusto and R.~Russo, ``{Superdescendants of the D1D5 CFT and their dual
  3-charge geometries},'' {\em JHEP} {\bf 1403} (2014) 007,
\href{http://arXiv.org/abs/1311.5536}{{\tt 1311.5536}}.

\bibitem{Maldacena:2015iua}
J.~Maldacena, D.~Simmons-Duffin, and A.~Zhiboedov, ``{Looking for a bulk
  point},'' {\em JHEP} {\bf 01} (2017) 013,
\href{http://arXiv.org/abs/1509.03612}{{\tt 1509.03612}}.

\end{thebibliography}
\end{document}